\documentclass[review]{elsarticle}

\usepackage{lineno,hyperref}
\modulolinenumbers[5]

\usepackage{times} 
\usepackage{amsmath} 
\usepackage{amssymb}  
\usepackage{mathtools}
\usepackage{subfig}
\usepackage{todonotes}
\usepackage{lscape}
\usepackage{cases}
\usepackage{enumitem}
\usepackage{changepage}

\begin{document}

\begin{frontmatter}

\title{An Affective Approach for Behavioral Performance Estimation and Induction   \tnoteref{mytitlenote}}
\tnotetext[mytitlenote]{This work is supported in part by National Science Foundation Awards IIS-1734272, CNS-1320561, and IIS-1208390.}

\author[mymainaddress]{Mustaffa Alfatlawi\corref{mycorrespondingauthor}}
\cortext[mycorrespondingauthor]{Corresponding author}
\ead{alfatlaw@msu.edu}

\address[mymainaddress]{Electrical and Computer Engineering Department, Michigan State University, East Lansing, MI 48824, USA}

\begin{abstract}
Emotions have a major interactive role in defining how humans interact with their environment by encoding their perception of external events and accordingly, influencing their cognition and decision-making process. Therefore, increasing attention has been directed toward integrating human affective states into system design in order to optimize the quality of task performance. In this work, we seize on the significant correlation between emotions and behavioral performance that is reported in several psychological studies and develop an online closed-loop design framework for Human-Robot Interaction (HRI). The proposed approach monitors the behavioral performance based on the levels of Pleasure, Arousal, and Dominance (PAD) states for the human operator and when required, applies an external stimulus which is selected to induce an improvement in performance. The framework is implemented on an HRI task involving a human operator teleoperating an articulated robotic manipulator. Our statistical analysis shows a significant decrease in pleasure, arousal, and dominance states as the behavioral performance deteriorates $(p < 0.05)$. Our closed-loop experiment that uses an audio stimulus to improve emotional state shows a significant improvement in the behavioral performance of certain subjects. 
\end{abstract}
\begin{keyword}
\texttt{\footnotesize{deep learning, Probabilistic models, Gaussian processes}}
\footnotesize{\MSC[2020] 68T07\sep  65C20 \sep 60G15}
\end{keyword}
\end{frontmatter}

\section{Introduction}
The worldwide competition to improve productivity, both quantitatively and qualitatively, has led to the deployment of robots at a rapid pace to play a supplementary role to humans in performing wide range of tasks in industry, transportation, and healthcare~\cite{hagele2016robots}. Predominantly, the existing human-robot integration methods maintain the central authority to human operators since robots are still lagging behind humans in their sensing and decision making capabilities~\cite{cummings2014man}. Even though the introduction of robots has been correlated to a boom in productivity~\cite{muro2015robots,graetz2015estimating}, it has contributed to a surge in the number of accidents, some of which have resulted in fatalities~\cite{engelberger2012robotics}. This public safety risk is, in part, attributed to the failure in accounting for the uncertainty in the behavioral performance of human operators, which is influenced by several factors such as motivation, workload, and fatigue, and may change during the course of task performance~\cite{national2007human}. In general, the deterioration in behavioral performance has contributed not only to historical tragedies, such as Three Mile Island power plant accident, Space Shuttle Challenger disaster, and Exxon Valdez oil spill, but has also contributed to day-to-day accidents in transportation, industry and health care systems~\cite{mitler1988catastrophes,reiner2012insidious,wheaton2014drowsy}.

Combating the uncertainty in behavioral performance encompasses two major challenges: first is the development of an online performance evaluation method with minimal interruption to the ongoing human-robot interaction, and second is the design of a mechanism which allows for external interventions to prevent the deteriorating behavioral performance.  
The evaluation of behavioral performance is typically conducted based on surveys and measurements collected before/after performing tasks. However, these methods, when applied to human-robot interaction systems, suffer from being disruptive to the human-robot integration and are incapable of differentiating between the error caused by behavioral performance and the error caused by the flaws in robot design~\cite{marathe2017confidence}. Thus, the focus was shifted toward developing more intuitive methods which can provide a genuine evaluation for behavioral performance. Notably, many studies have reported that behavioral performance can be inferred unobtrusively from the concurrent changes in behavioral and neurophysiological signals, such as facial and vocal expression, Electrocardiography (ECG), Electromyography (EMG), and Electroencephalography (EEG)~\cite{li2008measurement,murata2005evaluation,liu2018eeg,hu2017comparison,michail2008eeg,lin2010tonic}.
Authors  in \cite{huang2016eeg} develop a behavioral performance degradation detection and mitigation system, in which the EEG-based features are used to specify one of the two classes, fatigue and no-fatigue while driving a vehicle, and upon the detection of the fatigue class, a warning auditory signal is delivered to the participant in order to improve their degrading performance. A major drawback for utilizing neurophysiological-based features is that they may vary significantly from one person to another which makes it difficult to identify a generic signature for performance in-term of these features.

An abundance of psychological studies have investigated the bidirectional influence between the quality of performance and the concurrent emotions. For instance, a model of the relationship between the performance and emotions is presented in \cite{cai2011modeling}, which provides a quantitative description of the behavioral performance as a function of emotional state. Furthermore, the authors suggested that such relationship can be exploited to design a system that can estimate the emotion-induced performance and mitigate the performance degradation. In order to seize on the relationship between performance and emotions in HRI, measurements of emotions have to be accessible in real-time without any obstruction of the human-robot interaction. In this work, we utilize the so-called Pleasure-Arousal-Dominance (PAD) human emotional state model and estimate its state and ensuing quality of human performance from EEG signals. This approach has two major advantages: first, it provides an estimation for performance indexed by the PAD states in unobtrusive fashion, and second, it enables us to influence the performance, if necessary, using one of the emotions elicitation techniques available in literature. The major contributions of this work are:
 \begin{enumerate}
     \item \label{model}   We present a probabilistic model for the relationship between behavioral performance, emotions, and EEG signals that can be utilized to provide an online estimate of the performance of human operator without any interruption to the carried out activities.
     \item We leverage the above behavioral performance model to close the loop and design an intervention strategy, which utilizes an external emotional stimulus to maximize the probability that the behavioral performance is greater than a predefined safe value.  
 \end{enumerate}
 
The remainder of this paper is organized as follows. In Section~\ref{Theoretical Background}, we describe the problem studied in this paper and present some background material on the quantification of behavioral performance and emotion. In Section~\ref{Modeling of Behavioral Performance}, we present a probabilistic model of the behavioral performance in term of the emotional state. In Section~\ref{Behavioral Performance Closed Loop System}, we propose a closed-loop framework for monitoring and inducing behavioral performance in real-time. In Section~\ref{experimental setup}, we discuss the experimental setup and procedures to study the proposed framework, and discuss the obtained results in Section~\ref{results and discussion}. Finally, we conclude in Section~\ref{conclusion}.

\section{Theoretical Background}\label{Theoretical Background} 
In this section, we review the relationship between the behavioral performance and emotions from a psychological point of view. We then present the quantification methods that are adopted in this work to characterize emotions and behavioral performance. 

\subsection{Psychology for Emotions and Behavioral Performance}\label{psycho}
Quality of performance of a human operator is a volatile trait, that is prone to the level of skills, motivation, complexity of task, and several other parameters, and shows inconsistency during the course of performing a task~\cite{national2007human}. Nevertheless, a staggering amount of evidence indicates that there is an interplay between the quality of task performance and the existing mood {of the operator}. For instance, the perception of the quality of academic performance, is categorized as an emotional event because of the subsequent qualitative and quantitative emotional change~\cite{pekrun2009achievement}. A similar causal relationship between performance and emotions was observed in athletes with a significant correlation between the type of emotional status and the awareness of the possible outcomes of the match~\cite{seve2007performance}. Indeed, the emotional consequences of the perceived quality of performance is a well validated hypothesis and manipulation of performance evaluation is implemented as a technique to elicit certain types of emotions in laboratory environment~\cite{isen1993positive}.            
A large body of research in social psychology shows that the preexisting positive emotional status significantly improves the performance in creative tasks, and decreases the time required in decision making tasks~\cite{murphy2012oxford,fisher2004within}. In summary, the performance shows a transient response to the elicited emotions, while the perceived quality of performance instantaneously impacts the emotional status.

The behavioral performance is limited by the mental and physical capacities of the humans. Hence, from an engineering perspective, the behavioral performance is considered as an intrinsic property and is represented as a constraint for the overall performance of HRI systems. However, addressing the behavioral performance from a psychological point of view provides an alternative perspective in which the behavioral performance is partially extraneous due to the bidirectional causality between the performance and emotions. The advances in affective computing and the active research in emotion recognition and elicitation, set the stage to leverage the psychological relationship between behavioral performance and emotions toward control design for performance improvement. The research question addressed in this work can be stated as:

\begin{adjustwidth}{20pt}{20pt}
\textit{How to design a human-operated emotionally intelligent machine that is capable of :
\begin{enumerate}
    \item Evaluating the impact of the quality of behavioral performance on the operator's emotions.
    \item Managing the operator's emotions to improve the quality of performance and meet the safety standards.
\end{enumerate}} 
\end{adjustwidth}

\subsection{Quantification of Performance}
Task performance can be defined as the behavior exhibited by an operator as a part of the process that exploits available resources towards achieving a set of objectives. In general, three classes of task performance can be identified depending on the type and requirements of the task: routine, adaptive, and creative. Quantification of the quality of performance requires identifying measurable characteristics that collectively can present a numerical representation of the progress with respect to the goals. Thus, quantification of the performance relies solely on the nature of the task. 

In this work, we use the Quality of Teleoperation $\mathbf{QoT}$ which is a behavioral performance evaluation technique designed for robotic teleoperations~\cite{jia2014quality}. In particular, the behavioral performance of a human operator who is remotely commanding a robot to make it follow a certain trajectory can be numerically evaluated by calculating the $\mathbf{QoT}$ value using the following formula,
\begin{equation}
\mathbf{QoT} = \frac{1}{0.5 \sum_{i=1}^{2} \mathbf{p}_i},
\end{equation}
where $\mathbf{p}_1$ is the error rate defined by the number of instances of deviation from the trajectory per unit time, and $\mathbf{p}_2$ is the maximum distance of deviation from the trajectory.

\subsection{Quantification of Emotions}
There is an ongoing debate on the most effective method for emotion characterization. However, two models have been implemented extensively for emotion recognition, namely, the discrete and dimensional models. Discrete model, characterizes emotions into discrete states or groups of states. According to this model, Plutchik proposed eight basic emotions: anger, fear, sadness, happiness, surprise, disgust, acceptance, and anticipation~\cite{plutchik2003emotions}. On the other hand, the dimensional model,  suggests that the emotion can be characterized  by three dimensional space spanned by pleasure, arousal, and dominance vectors; all other emotions can be represented as a linear combination of these three emotions~\cite{russell1977evidence,russell1980circumplex}. The continuous scale for emotions provided by the dimensional model makes it preferable when emotion recognition problem is approached using regression-based estimation methods, which is the case in this work. 

Emotion recognition is an active research area in affective computing which aims to develop methods for assessing the emotional status by identifying its impact on the behavioral and physiological signals. Among all investigated signals for the study of emotional impact, EEG is considered to be a reliable source of information pertaining to the emotional changes, due to its proximity to the origin of emotional response and being less prone to individual differences and disingenuity. The process of EEG acquisition and analysis is relatively more intrusive and technically demanding in comparison with other physiological signals, due to the number of electrodes and data storage capacity required for EEG recording in order to achieve desired fidelity in the outcomes of emotion recognition~\cite{muhl2014survey}. Nevertheless, EEG has been investigated rigorously in an attempt to identify distinctive neurophysiological signatures for different emotions. Numerous studies have shown a significant correlation between emotions and certain characteristics of EEG signals and its frequency bands: delta (0.5-4 Hz), theta (3-7Hz), alpha (8-13 Hz), beta (14-29 Hz), and gamma (30-47 Hz). These characteristics can be classified into time-domain, frequency-domain, and time-frequency domain characteristics. 
\section{Modeling of Behavioral Performance}\label{Modeling of Behavioral Performance}
In this section, we present a probabilistic interpretation of the psychological relationships between the behavioral performance, PAD states, and EEG signals, presented in Section ~\ref{Theoretical Background}. Also, we describe the learning techniques and the required experimental datasets needed to obtain a generative model that can be utilized to provide real-time prediction of the behavioral performance.  

\subsection{Problem Formulation}
Consider an HRI application in which the human operator is controlling a remote robot via visual feedback to perform a set of tasks. Meanwhile, the EEG data is collected by a wearable headset. Let the behavioral performance and EEG-based features be the observed variables denoted by $q\in \mathbb{R}$ and $\mathbf{e}\in \mathbb{R}^n$, respectively. Let the PAD states be the latent variable denoted as $\mathbf{f}\in \mathbb{R}^3$.  
We work under the hypothesis that for a given task, the variation in behavioral performance is determined by the PAD state. Additionally, we assume that the EEG-based features that measure the behavioral performance are also determined by the PAD state and accordingly, we model    \begin{align}
        \mathbf{e}&=\mathfrak{g}_e(\mathbf{f})+ \epsilon_f \label{EEG process}\\ 
         q&=\mathfrak{g}_b(\mathbf{f})+ \epsilon_b,  \label{performance process}
    \end{align}
where the observation of $q$ in equation \eqref{performance process} is an unknown function of the PAD state $\mathfrak{g}_b$ contaminated with the measurement noise $\epsilon_b \backsim \mathcal{N}(0,\sigma_b^2)$, and the observations of $\mathbf{e}$ in equation \eqref{EEG process} is an unknown function of the PAD state $\mathfrak{g}_e$ contaminated with the measurement noise $\epsilon_f \backsim \mathcal{N}(0,\sigma_f^2)$.
Our objective is to infer the probability distribution of behavioral performance conditioned on the EEG-based features, i.e.,  $q\backsim \mathbf{p}(q|\mathbf{e})$.

\subsection{Inference of Behavioral Performance}\label{Inference of Behavioral Performance}
\begin{figure}[!t]
\centering
\subfloat[Performance Induction Graphical Model]
{
\label{graphical_model_perf}
\includegraphics[width=2in]{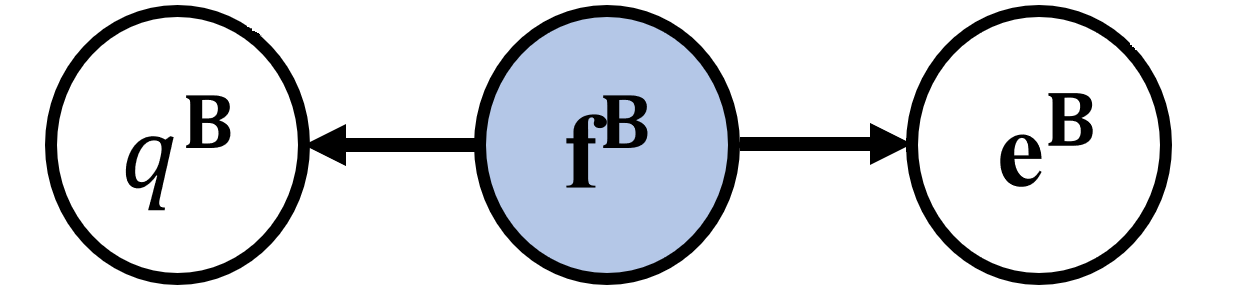}
}
\subfloat[Emotion Elicitation  Graphical Model]
{
\label{graphical_model_emo}
\includegraphics[width=2in]{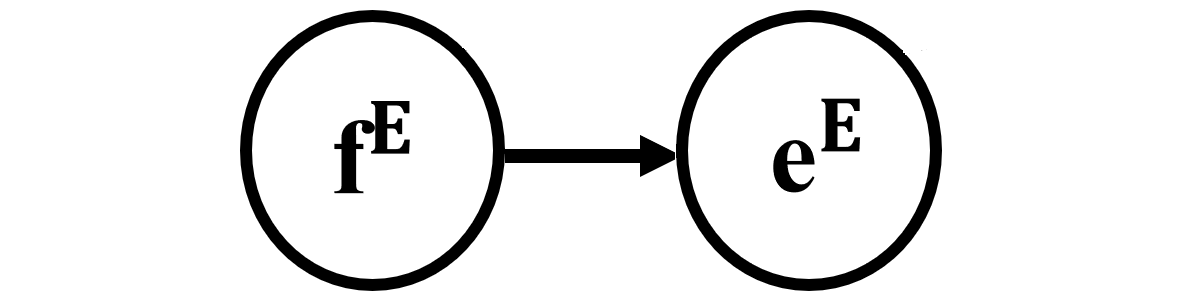}
}
\caption{Behavioral performance process graphical model for the probabilistic interaction between behavioral performance ${q}^{\text{B}}$, PAD states $\mathbf{f}^{\text{B}}$, and EEG-based features $\mathbf{e}^{\text{B}}$ (lower panel), and emotion elicitation process graphical model for the probabilistic interaction between, PAD states $\mathbf{f}^{\text{E}}$, and PAD states $\mathbf{e}^{\text{E}}$ for a behavioral performance process (lower panel).}
\end{figure}

Consider a behavioral performance-induction experiment in which a human operator tracks a predefined trajectory. Suppose  the behavioral performance and the concurrent EEG data are recorded during the experiment. Consider the dataset $\{\mathbf{q}^{\text{B}},\mathbf{E}^{\text{B}}\}$, where $\mathbf{E}^{\text{B}}=\begin{bmatrix}\mathbf{e}^{\text{B}}_1 & \cdots & \mathbf{e}^{\text{B}}_{m_b} \end{bmatrix}$ is a matrix whose $m_b$ columns are observations of EEG-based features $\mathbf{e}^{\text{B}}_i \in \mathbb{R}^{n}$ stimulated by the visual perception of performance, and $\mathbf{q}^{\text{B}}=\begin{bmatrix}q^{\text{B}}_1 & \cdots & q^{\text{B}}_{m_b} \end{bmatrix}$ is a column vector whose $m_b$ entries are observations of behavioral performance $q^{\text{B}}_i \in \mathbb{R}$.
The graphical model in Figure~\ref{graphical_model_perf} represents the interaction between behavioral performance ${q}^{\mathbf{\text{B}}}$, PAD states $\mathbf{f}^{\mathbf{\text{B}}}$, and EEG-based feature $\mathbf{e}^{\mathbf{\text{B}}}$.
The joint distribution for behavioral performance and PAD states can be expressed according to the graphical model in Figure~\ref{graphical_model_perf} as
\begin{align}\label{joint_dist0}
     \mathbf{p}(q^{\text{B}},\mathbf{f}^{\text{B}}|\mathbf{e}^{\text{B}})
     = \mathbf{p}(q^{\text{B}}|\mathbf{f}^{\text{B}}) \mathbf{p}(\mathbf{e}^{\text{B}}|\mathbf{f}^{\text{B}}) \mathbf{p}(\mathbf{f}^{\text{B}}) \frac{1}{\mathbf{p}(\mathbf{e}^{\text{B}})}
     = \mathbf{p}(q^{\text{B}}|\mathbf{f}^{\text{B}})  \mathbf{p}(\mathbf{f}^{\text{B}}|\mathbf{e}^{\text{B}}).
 \end{align}
The major obstacle to learning distribution in~\eqref{joint_dist0} from the dataset $=\{\mathbf{q}^{\text{B}},\mathbf{E}^{\text{B}}\}$, is the latent variable distribution  $\mathbf{p}(\mathbf{f}^{\text{B}})$.
 \\In order to overcome this obstacle, we conduct another experiment, namely emotion elicitation experiment, that results in triggering different emotions for a human participant through the delivery of visual stimuli each of which is labeled with the resulting of PAD states, which illustrates the interaction between the elicited PAD states $\mathbf{f}^{\text{E}}$ and the corresponding EEG-based features $\mathbf{e}^{\text{E}}$ during the emotion elicitation experiment (see  graphical model in Figure~\ref{graphical_model_emo}).

Suppose that we are given the dataset $\{\mathbf{F}^{\text{E}},\mathbf{E}^{\text{E}}\}$, where $\mathbf{E}^{\text{E}}=\begin{bmatrix}\mathbf{e}^{\text{E}}_1 & \cdots & \mathbf{e}^{\text{E}}_{m_f} \end{bmatrix}$ is a matrix whose $m_f$ columns are observations of EEG-based features $\mathbf{e}^{\text{E}}_i \in \mathbb{R}^{n}$ under different visual emotional stimuli, and $\mathbf{F}^{\text{E}}=\begin{bmatrix}\mathbf{f}^{\text{E}}_1 & \cdots & \mathbf{f}^{\text{E}}_{m_f} \end{bmatrix}$ is a matrix whose $m_f$ columns are labeled PAD states $\mathbf{f}^{\text{E}}_i =\begin{bmatrix} p_i & a_i & d_i \end{bmatrix}\in \mathbb{R}^{3}$. Then, the labeled dataset $\{\mathbf{F}^{\text{E}},\mathbf{E}^{\text{E}}\}$ can be utilized to infer the posterior distribution $\mathbf{p}(\mathbf{f}^{\text{B}}|\mathbf{e}^{\text{B}})$ for every $\mathbf{e}^{\text{B}} \in \{ \mathbf{e}^{\text{B}}_1 , \cdots , \mathbf{e}^{\text{B}}_{m_b}\}$. Furthermore, the marginal distribution~$\mathbf{p}(q^{\text{B}}|\mathbf{e}^{\text{B}})$ can be obtained from~\eqref{joint_dist0} by the marginal likelihood,
\begin{align}
    \mathbf{p}(q^{\text{B}}|\mathbf{e}^{\text{B}})&= \int \mathbf{p}(q^{\text{B}}|{\mathbf{f}}^{\text{B}}) \mathbf{p}({\mathbf{f}}^{\text{B}}|\mathbf{e}^{\text{B}})~d{\mathbf{f}}^{\text{B}}. \label{joint_dist5}
 \end{align} 
 
In order to avoid the computational cost of integration, Laplace approximation~\cite{butler2007saddlepoint} can be used to approximate the marginal likelihood in~\eqref{joint_dist5} by,
\begin{align} \label{marginal_likelihood2}
    \mathbf{p}(q^{\text{B}}|\mathbf{e}^{\text{B}})& \approx \mathbf{p}(q^{\text{B}}|\hat{\mathbf{f}}^{\text{B}}) \mathbf{p}(\hat{\mathbf{f}}^{\text{B}}|\mathbf{e}^{\text{B}}) (2 \pi)^{\frac{3}{2}} {\vert \Sigma_f \vert}^{\frac{1}{2}},  
 \end{align} 
where $\displaystyle \hat{\mathbf{f}}^{\text{B}} =     \underset{{\mathbf{f}}^{\text{B}}}{\text{arg}}~\text{max}~~  \mathbf{p}({\mathbf{f}}^{\text{B}}|\mathbf{e}^{\text{B}})$ and $\displaystyle \Sigma_f =     \left(- \nabla^2 ~\text{log}\left(\mathbf{p}({\mathbf{f}}^{\text{B}}|\mathbf{e}^{\text{B}})\right)\right)^{-1} |_{{\mathbf{f}}^{\text{B}}=\hat{\mathbf{f}}^{\text{B}}}$.

\begin{figure}[!t]
\centering
\includegraphics[width=3in]{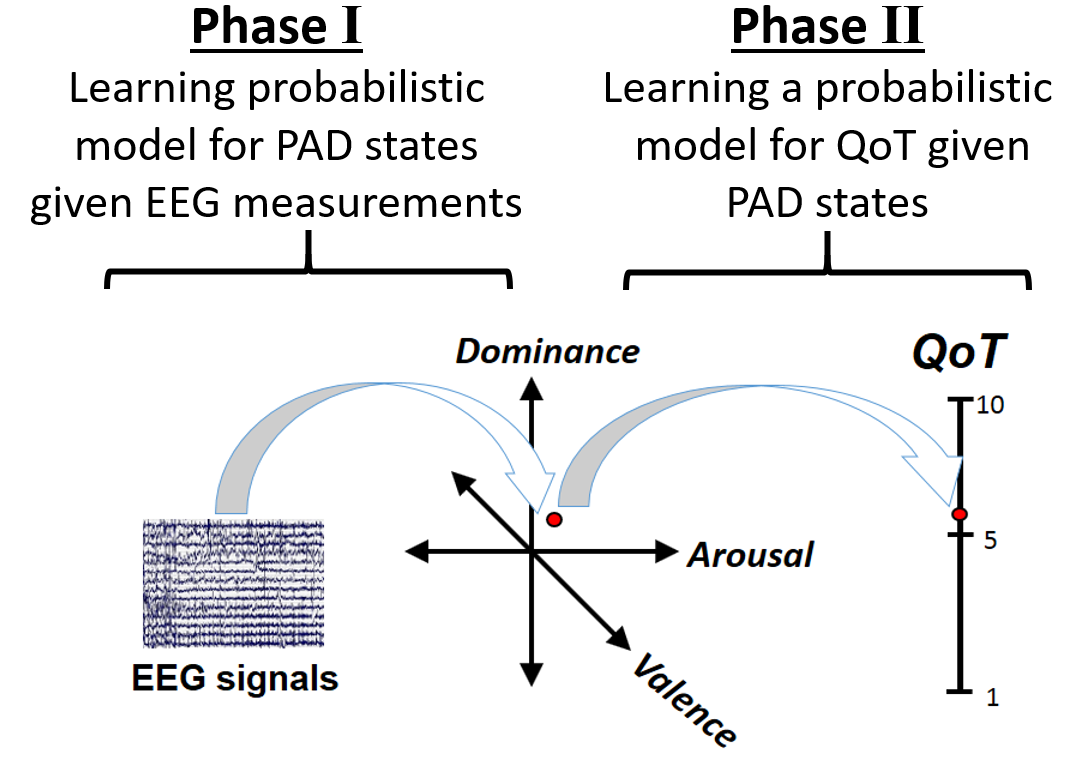}
\caption{The two phases for learning a probabilistic regression model for behavioral performance: the first phase is for learning the conditional distribution of PAD states given the EEG measurements, and the second phase is for learning the conditional distribution of the $\mathbf{QoT}$ given the PAD states}
\label{overall}
\end{figure}

Now, we use the posterior distribution in~\eqref{marginal_likelihood2} to make inference about behavioral performance as depicted in Figure~\ref{overall}, and described below.

\noindent 
\textbf{Phase I:}  Consider two datasets of observations for the EEG-based features $\mathbf{E}^{\text{E}}=\begin{bmatrix} \mathbf{e}^{\text{E}}_1 & \cdots & \mathbf{e}^{\text{E}}_{m_f} \end{bmatrix}\in \mathbf{R}^{n \times {m_f}}$, and the PAD state labels $\mathbf{F}^{\text{E}}=\begin{bmatrix} \mathbf{\hat f}^{\text{E}}_1 & \cdots & \mathbf{\hat f}^{\text{E}}_{m_f} \end{bmatrix} \in \mathbf{R}^{3 \times {m_f}}$ associated with the corresponding stimuli.
Let the PAD state $\mathbf f$ be an unknown function of the EEG-based features $\mathbf{e}$ and be defined by $\mathbf{f}=\mathbf{g}_f(\mathbf{e})$. Additionally, let the labels $\mathbf{\hat f}^{\text{E}}_i$ be noisy estimates of the true PAD state $\mathbf{g}_f(\mathbf{e}_i^{\text{E}})$ and be defined by  
\begin{equation}
    \mathbf{\hat f}^{\text{E}}_i=\mathbf{g}_f(\mathbf{e}_i^{\text{E}})+ \epsilon_f,  \nonumber
\end{equation}
where 
$\epsilon_f\backsim \mathcal{N}(0,\sigma_f^2)$ is the measurement noise. Given a new observation for EEG-based features $\mathbf{e}^{\text{B}}_*$, our objective, depicted as Phase I in Figure~\ref{overall}, is to infer the predictive distribution of the PAD state $\mathbf{f}^{\text{B}}_{*}=\mathbf{g}_f(\mathbf{e}^{\text{B}}_*)$, i.e., the posterior distribution $\mathbf{p}(\mathbf{f}^{\text{B}}_{*}|\mathbf{e}^{\text{B}}_*)$. 

\begin{figure}[!b]
\centering
\includegraphics[width=4in] {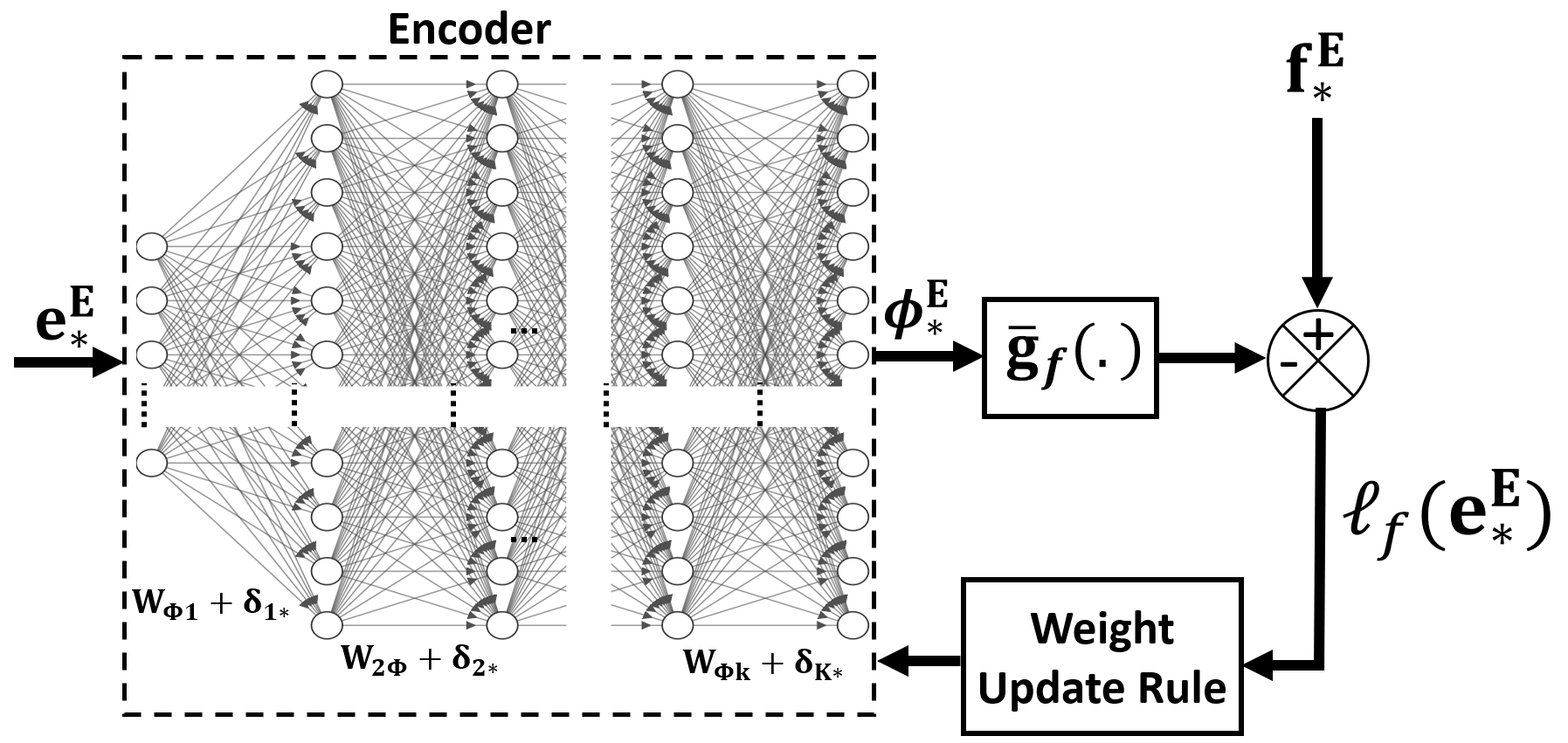}
\caption{The update of weights for K-layers neural network using Backpropagation method}
\label{NN_BP}
\end{figure}

Let each entry of $\mathbf{g}_f(\mathbf{e}) \in \mathbb{R}^3$ be a be a realization of a zero-mean Gaussian process  with the covariance function $k_f^\ell(\mathbf{e},\acute{\mathbf{e}})$, for $\ell \in \{1,2, 3\}$, i.e., $ (\mathbf{g}_f(\mathbf{e}))_{\ell} \backsim \text{GP}(\mathbf{0},k_f^\ell(\mathbf{e},\acute{\mathbf{e}}))$, where $\mathbf{e}$ and $\acute{\mathbf{e}}$ are two vectors of EEG-based features. A key factor for accurate representation as a Gaussian process is the choice of the kernel function $k_f^\ell$ which represents our assumption about the function of interest $\mathbf{g}_f$. The kernel functions $k_f^\ell(\mathbf{e},\acute{\mathbf{e}})$ are learnt using Deep Belief Network (DBN) in two steps following the approach proposed in~\cite{hinton2008using}: 

First, a DBN is learned from unlabeled dataset $\mathbf{E}^{\text{B}}$ in an unsupervised manner. The details of the training processes are discussed in Section~\ref{results and discussion}. The resulting DBN  provides a nonlinear map $\mathbf{\Phi}:\mathcal{E} \rightarrow \mathcal{Z}$ from the space of EEG-based features $\mathcal{E} \subset \mathcal{R}^n$ into a higher dimensional space of latent features $\mathcal{Z} \subset \mathcal{R}^N$, parameterized by the weights $\mathbf{W}_{\phi}=\{\mathbf{W}_{\phi 1},\cdots,\mathbf{W}_{\phi K}\}$ of the $K$-layer DBN, as it shown in Figure~\ref{NN_BP}, where $N \gg n$. The main advantage of the DBN is that the features $\boldsymbol{\phi} \in \mathcal{Z}$ encode the higher-order correlations of the input EEG-based features $\mathbf{e} \in \mathcal{E}$, which makes the application of covariance function in term of encoded features $\boldsymbol{\phi}= \Phi(\mathbf{e}|\mathbf{W}_{\phi})$ more efficient than the application using the input EEG-based features $\mathbf{e}$ {~\cite{hinton2008using}}. As a result, for each pair of EEG-based features $\{ \mathbf{e}_i,\mathbf{e}_j \}$, the covariance function is expressed in term of their corresponding latent features $\boldsymbol {\phi}_i=\Phi(\mathbf{e}_i|\mathbf{W}_{\phi})$ and $\boldsymbol{\phi}_j=\Phi(\mathbf{e}_j|\mathbf{W}_{\phi})$, is written as
\begin{align} \label{covariance function_latent}
      k_f^\ell(\boldsymbol{\phi}_i,\boldsymbol{\phi}_j)=\alpha_{\phi}^\ell \exp{(-\frac{1}{2 \beta_{\phi}^\ell}\left\|\boldsymbol{\phi}_i-\boldsymbol{\phi}_j\right\|^2)}, \; \ell \in \{1,2,3\},  \nonumber
\end{align}
where $\alpha_{\phi}^\ell$ and $\beta_{\phi}^\ell$ are the variance and length scale of the latent features, respectively. 

Second, the weights $\mathbf{W}_{\phi}$ of the DBN and the hyperparameters $\alpha_{\phi}^\ell$ and $\beta_{\phi}^\ell$   can be further tuned from the labeled dataset $\{\mathbf{F}^{\text{E}},\mathbf{E}^{\text{E}}\}$ in a supervised manner as shown in Figure~\ref{NN_BP}. Thus, for each $\mathbf{e}^{\text{E}}_* \in \{\mathbf{e}^{\text{E}}_1, \cdots, \mathbf{e}^{\text{E}}_{m_f}\}$ and its corresponding $\mathbf{f}^{\text{E}}_* \in \{\mathbf{f}^{\text{E}}_1, \cdots, \mathbf{f}^{\text{E}}_{m_f}\}$, backpropagation method {~\cite{haykin2009neural}} is implemented to update $\{\mathbf{W}_{\phi}, \alpha_{\phi},\beta_{\phi}\}$ such that the loss function
\begin{align}
    l_f(\mathbf{e}^{\text{E}}_*)= \left\|\mathbf{f}^{\text{E}}_*-\bar{\mathbf{g}}_f(\mathbf{e}^{\text{E}}_*)\right\|^2, 
\end{align}
is minimized.

Given the training datasets $\mathbf{E}^{\text{E}}$ and  $\mathbf{F}^{\text{E}}$, a new vector of EEG-based features $\mathbf{e}^B_*$ with associated PAD state $\mathbf{f}^B_*$, and the latent features for the training dataset $\mathbf{E}^{\text{E}}$ denoted as $\boldsymbol{\Phi}^{\text{E}}= \begin{bsmallmatrix} \boldsymbol{\phi}_1^{\text{E}} & \cdots & \boldsymbol{\phi}_{m_f}^{\text{E}}\end{bsmallmatrix}=\begin{bsmallmatrix}\Phi(\mathbf{e}_1^{\text{E}}|\mathbf{W}_{\phi}) & \cdots & \Phi(\mathbf{e}_{mf}^{\text{E}}|\mathbf{W}_{\phi})\end{bsmallmatrix}\in \mathcal{R}^{N \times mf}$ and for the new EEG-based features vector $\mathbf{e}^B_*$ denoted as $\boldsymbol{\phi}_*^{\text{B}}=\Phi(\mathbf{e}^{\text{B}}_*|\mathbf{W}_{\phi}) \in \mathcal{R}$, the joint distribution of $(\mathbf{\hat f}_1^E)_\ell, \ldots, (\mathbf{\hat f}_{m_f}^E)_\ell, (\mathbf{f}^B_*)_\ell$, for $\ell \in \{1,2,3\}$, is
\begin{equation}\label{mean_cov}
    \mathbf{p}((\mathbf{\hat f}_1^E)_\ell, \ldots, (\mathbf{\hat f}_{m_f}^E)_\ell, (\mathbf{f}^B_*)_\ell
    = \mathcal{N} \left(0,\begin{bsmallmatrix} \mathbf{K}^{\text{E}}_\ell(\boldsymbol{\phi}^{\text{E}},\boldsymbol{\phi}^{\text{E}})+\sigma_f^2\mathbf{I}_{m_f} & \mathbf{K}^{\text{E}}_\ell (\boldsymbol{\phi}^{\text{E}},\boldsymbol{\phi}_*^{\text{B}}) \\ \mathbf{K}^{\text{E}}_\ell (\boldsymbol{\phi}_*^{\text{B}},\boldsymbol{\phi}^{\text{E}}) & k_f^\ell(\boldsymbol{\phi}^{\text{B}}_*,\boldsymbol{\phi}^{\text{B}}_*)\end{bsmallmatrix}\right), \nonumber
\end{equation}
where $\mathbf{I}_{m_f}$ is $m_f \times m_f$ identity matrix, and the entries of covariance matrix are calculated using the covariance formula in~\eqref{covariance function_latent} as follows
\begin{eqnarray}
  \mathbf{K}^{\text{E}}_\ell(\boldsymbol{\phi}^{\text{E}},\boldsymbol{\phi}^{\text{E}})=
  \begin{bsmallmatrix}
    k_f^\ell(\boldsymbol{\phi}^{\text{E}}_1,\boldsymbol{\phi}^{\text{E}}_1) & \ldots & k_f^\ell(\boldsymbol{\phi}^{\text{E}}_1,\boldsymbol{\phi}^{\text{E}}_{m_f})\\
    \vdots & \ddots & \vdots \\
    k_f^\ell(\boldsymbol{\phi}^{\text{E}}_{m_f},\boldsymbol{\phi}^{\text{E}}_1) & \ldots & k_f^\ell(\boldsymbol{\phi}^{\text{E}}_{m_f},\boldsymbol{\phi}^{\text{E}}_{m_f})
  \end{bsmallmatrix}, & \mathbf{K}^{\text{E}}_\ell(\boldsymbol{\phi}^{\text{E}},\phi_*^{\text{B}})={\mathbf{K}^{\text{E}}_\ell(\phi_*^{\text{B}},\boldsymbol{\phi}^{\text{E}})}^\top=\begin{bsmallmatrix}
    k_f^\ell(\boldsymbol{\phi}^{\text{E}}_1,\boldsymbol{\phi}_*^{\text{B}}) & \cdots & k_f^\ell(\phi^{\text{E}}_{m_f},\boldsymbol{\phi}_*^{\text{B}}) 
    \end{bsmallmatrix}.  \nonumber  
\end{eqnarray}

Hence, given a new input measurement $\mathbf{e}^{\text{B}}_{*}$, the predicted PAD states is defined by the distribution, $\mathbf{f}^{\text{B}}_{*}|(\mathbf{e}^{\text{B}}_{*},\mathbf{E}^{\text{E}}, \mathbf{F}^{\text{E}}) \backsim \mathcal{N}(\bar{\mathbf{g}}_f(\mathbf{e}^{\text{B}}_{*}), \bar{\mathbf{K}}_{f}(\mathbf{e}^{\text{B}}_{*}))$, with mean and covariance functions defined by $\bar{\mathbf{g}}_f(\mathbf{e}^{\text{B}}_*) = [  (\bar{\mathbf{g}}_f(\mathbf{e}^{\text{B}}_*))_1,   (\bar{\mathbf{g}}_f(\mathbf{e}^{\text{B}}_*))_2,   (\bar{\mathbf{g}}_f(\mathbf{e}^{\text{B}}_*))_3]^\top$, and $   \bar{\mathbf{K}}^f(\mathbf{e}^{\text{B}}_*) = \text{diag}(   \bar{\mathbf{K}}^f_1(\mathbf{e}^{\text{B}}_*),    \bar{\mathbf{K}}^f_2(\mathbf{e}^{\text{B}}_*),    \bar{\mathbf{K}}^f_3(\mathbf{e}^{\text{B}}_*))$, respectively,  where 
\begin{equation}\label{pdf_PAD}
\begin{aligned}
    (\bar{\mathbf{g}}_f(\mathbf{e}^{\text{B}}_*))_\ell &=\mathbf{K}^{\text{E}}(\boldsymbol{\phi}^{\text{B}}_*, \mathbf{\Phi}^{\text{E}})[\mathbf{K}_\ell ^{\text{E}}(\mathbf{\Phi}^{\text{E}},\mathbf{\Phi}^{\text{E}})+ \sigma_n^2\mathbf{I}_{m_f}]^{-1}\mathbf{F}^{\text{E}}\\ 
    \bar{\mathbf{K}}^f_\ell(\mathbf{e}^{\text{B}}_*)&=\mathbf{K}^{\text{E}}_\ell(\boldsymbol{\phi}^{\text{B}}_*,\boldsymbol{\phi}^{\text{B}}_*)-\mathbf{K}^{\text{E}}_\ell(\boldsymbol{\phi}^{\text{B}}_*,\mathbf{\Phi}^{\text{E}})[\mathbf{K}^{\text{E}}_\ell(\mathbf{\Phi}^{\text{E}},\mathbf{\Phi}^{\text{E}})+ \sigma_n^2\mathbf{I}_{m_f}]^{-1} \mathbf{K}^{\text{E}}_\ell(\mathbf{\Phi}^{\text{E}},\boldsymbol{\phi}^{\text{B}}_*), \; \ell \in \{1,2,3\},
\end{aligned}
\end{equation}.


\noindent 
\textbf{Phase II:}  Given two datasets of observations for EEG-based features $\mathbf{E}^{\text{B}}=\begin{bmatrix} \mathbf{e}^{\text{B}}_1 & \cdots & \mathbf{e}^{\text{B}}_{m_b} \end{bmatrix}\in \mathbf{R}^{n \times {m_b}}$ and behavioral performance $\mathbf{q}^{\text{B}}=\begin{bmatrix} q^{\text{B}}_1 & \cdots & q^{\text{B}}_{m_b} \end{bmatrix} \in \mathbf{R}^{m_b}$, with the underlying model,
   \begin{align}
    {q}_i^{\text{B}}&=g_b(\mathbf{f}^{\text{B}}_{i})+ \epsilon_q, \nonumber
   \end{align} 
   where $\mathbf{f}^{\text{B}}_{i}$ is the PAD states vector with the posterior distribution $\mathbf{f}^{\text{B}}_{i}|(\mathbf{e}^{\text{B}}_{*},\mathbf{E}^{\text{E}}, \mathbf{F}^{\text{E}}) \backsim \mathcal{N}\left(\bar{\mathbf{g}}_f \left(\mathbf{e}_{i}^{\text{B}}\right), \bar{\mathbf{K}}_f\left(\mathbf{e}_{i}^{\text{B}}\right)\right)$, where the mean and covariance functions are defined in \eqref{pdf_PAD}, and the observed $q_i^{\text{B}}$ is an unknown function $g_b(\mathbf{f}^{\text{B}}_{i})$ contaminated with the measurement noise $\epsilon_q\backsim \mathcal{N}(0,\sigma_b^2)$. The problem, which is depicted as Phase II in Figure~\ref{overall}, is to infer the posterior $\mathbf{p}(q^{\text{B}}_{*}| (\mathbf{e}^{\text{B}}_{*},\mathbf{E}^{\text{E}}, \mathbf{F}^{\text{E}}))$ that defines the distribution for the predicted behavioral performance $q^{\text{B}}_{*}=g_b(\mathbf{f}^{\text{B}}_{*})$.
    The posterior $\mathbf{p}(q^{\text{B}}_{*}|\mathbf{f}^{\text{B}}_*)$ is approximated using equation~\eqref{marginal_likelihood2}, wherein $\hat{\mathbf{f}}^{\text{B}}_{*}=\bar{\mathbf{g}}_f(\mathbf{e}_*^{\text{B}})$, since $\mathbf{f}^{\text{B}}_{i}|(\mathbf{e}^{\text{B}}_{*},\mathbf{E}^{\text{E}}, \mathbf{F}^{\text{E}})$ is modeled as a Gaussian distribution. 
Hence, the problem reduces to the inference of the posterior $\mathbf{p}(q^{\text{B}}_{*}|\bar{\mathbf{g}}_f(\mathbf{e}_*^{\text{B}}))$ that defines the distribution for the predicted $q^{\text{B}}_{*}$ given the mean function $\bar{\mathbf{g}}_f(\mathbf{e}_*^{\text{B}})$ of the new observation for EEG-based features $\mathbf{e}^{\text{B}}_*$.

   Let ${g}_b(\bar{\mathbf{g}}_f(\mathbf{e}))\backsim\text{GP}(\mathbf{0},k_b(\bar{\mathbf{g}}_f(\mathbf{e}),\bar{\mathbf{g}}_f(\acute{\mathbf{e}}))$ be a zero-mean Gaussian process  with a covariance function represented by $k_b\left(\bar{\mathbf{g}}_f(\mathbf{e}),\bar{\mathbf{g}}_f(\acute{\mathbf{e}})\right)$. Let ${{\mathbf{F}}}^{\text{B}} \triangleq \begin{bmatrix} \bar{\mathbf{f}}^{\text{B}}_1 & \cdots & \bar{\mathbf{f}}^{\text{B}}_{m_b} \end{bmatrix} \in \mathbf{R}^{3 \times {m_b}}$, where $\bar{\mathbf{f}}^{\text{B}}_i \triangleq \bar{\mathbf{g}}_f(\mathbf{e}_i^{\text{B}})$ is defined in~\eqref{mean_cov}. We select the following covariance function
    \begin{equation}\label{performance covariance function}
      k_b(\bar{\mathbf{f}}^{\text{B}}_i,\bar{\mathbf{f}}^{\text{B}}_j)=\alpha_{g}\exp{(-\frac{1}{2 \beta_{g}}\left\|\bar{\mathbf{f}}^{\text{B}}_i-\bar{\mathbf{f}}^{\text{B}}_j\right\|^2)},
\end{equation}
and $\alpha_{g}$ and $\beta_{g}$ are the variance and length scale, respectively. The hyperparameters $\alpha_{g}$ and $\beta_{g}$, and measurement noise $\sigma_b^2$ are learned using datasets  $\mathbf{F}^{\text{B}}$ and $\mathbf{q}^{\text{B}}$; see Section~\ref{results and discussion} for details of the process.

  Let $\bar{\mathbf{f}}^{\text{B}}_* \triangleq \bar{\mathbf{g}}_f(\mathbf{e}_*^{\text{B}})$ for a new observations for EEG-based features $\mathbf{e}_*^{\text{B}}$. 
   Then, the behavioral performance states have the joint Gaussian distribution
   \begin{equation}
    \mathbf{p}({g}_b(\bar{\mathbf{f}}^{\text{B}}_1), \dots ,{g}_b(\bar{\mathbf{f}}^{\text{B}}_{m_b}),{g}_b(\bar{\mathbf{f}}^{\text{B}}_*))= \mathcal{N}\left(0,\begin{bsmallmatrix} \mathbf{K}^{\text{B}}(\mathbf{F}^{\text{B}},\mathbf{F}^{\text{B}})+\sigma_b^2\mathbf{I}_{m_b} & \mathbf{K}^{\text{B}}(\mathbf{F}^{\text{B}},\bar{\mathbf{f}}^{\text{B}}_*) \\ \mathbf{K}^{\text{B}}(\bar{\mathbf{f}}^{\text{B}}_*,\mathbf{F}^{\text{B}}) & k_b(\bar{\mathbf{f}}^{\text{B}}_*,\bar{\mathbf{f}}^{\text{B}}_*)\end{bsmallmatrix}\right), \nonumber
\end{equation}
 where $\mathbf{I}_{m_b}$ is $m_b \times m_b$ identity matrix, and the entries of covariance matrix $\mathbf{K}^{\text{B}}_{i,j}= k_b(\bar{\mathbf{f}}^{\text{B}}_i,\bar{\mathbf{f}}^{\text{B}}_j)$ for $i,j \in \{1,\dots,m_b, *\}$, 
 and the entries of covariance matrix are calculated using the covariance function in~\eqref{performance covariance function} as follows
\begin{eqnarray}
  \mathbf{K}^{\text{B}}(\mathbf{F}^{\text{B}},\mathbf{F}^{\text{B}})=
  \begin{bsmallmatrix}
    k_b(\bar{\mathbf{f}}^{\text{B}}_1,\bar{\mathbf{f}}^{\text{B}}_1) & \ldots & k_b(\bar{\mathbf{f}}^{\text{B}}_1,\bar{\mathbf{f}}^{\text{B}}_{m_b})\\
    \vdots & \ddots & \vdots \\
    k_b(\bar{\mathbf{f}}^{\text{B}}_{m_b},\bar{\mathbf{f}}^{\text{B}}_1) & \ldots & k_b(\bar{\mathbf{f}}^{\text{B}}_{m_b},\bar{\mathbf{f}}^{\text{B}}_{m_b})
  \end{bsmallmatrix}, & \mathbf{K}^{\text{B}}(\mathbf{F}^{\text{B}},\bar{\mathbf{f}}^{\text{B}}_*)={\mathbf{K}^{\text{B}}(\bar{\mathbf{f}}^{\text{B}}_*,\mathbf{F}^{\text{B}})}^\top=\begin{bsmallmatrix}
    k_b(\bar{\mathbf{f}}^{\text{B}}_1,\bar{\mathbf{f}}^{\text{B}}_*) & \cdots & k_b(\bar{\mathbf{f}}^{\text{B}}_{m_b},\bar{\mathbf{f}}^{\text{B}}_*) 
    \end{bsmallmatrix}.  \nonumber  
\end{eqnarray}

Finally, given a new observation for the EEG-based features $\mathbf{e}^{\text{B}}_{*}$, the predicted noise-free behavioral performance is defined by the distribution, $q^{\text{B}}_{*}|({\mathbf{e}}^{\text{B}}_*,\mathbf{E}^{\text{E}}, \mathbf{F}^{\text{E}},\mathbf{F}^{\text{B}}, \mathbf{q}^{\text{B}}) \backsim \mathcal{N}\left(\bar{{g}}_b(\bar{\mathbf{e}}^{\text{B}}_*), \bar{k}_{b}(\bar{\mathbf{f}}^{\text{B}}_*)\right)$, with mean and variance functions defined by
\begin{equation}\label{behavioral_perf}
\begin{aligned}
    \bar{{g}}_b(\bar{\mathbf{e}}^{\text{B}}_*)&=\left(\bar{g}_b \circ \bar{\mathbf{g}}_f\right)(\mathbf{e}^{\text{B}}_{*})= \mathbf{K}^{\text{B}}(\bar{\mathbf{f}}^{\text{B}}_*,\mathbf{F}^{\text{B}}) \left[\mathbf{K}^{\text{B}}(\mathbf{F}^{\text{B}},\mathbf{F}^{\text{B}})+ \sigma_b^2\mathbf{I}_{m_b} \right]^{-1}\mathbf{q}^{\text{B}}\\ 
    \bar{k}_{b}(\bar{\mathbf{e}}^{\text{B}}_*)&= \left(\bar{k}_{b}\circ \bar{\mathbf{g}}_f\right)(\mathbf{e}^{\text{B}}_{*})=\mathrm{k}^{\text{B}}(\bar{\mathbf{f}}^{\text{B}}_*,\bar{\mathbf{f}}^{\text{B}}_*)-\mathbf{K}^{\text{B}}(\bar{\mathbf{f}}^{\text{B}}_*,\mathbf{F}^{\text{B}}) \left[\mathbf{K}^{\text{B}}(\mathbf{F}^{\text{B}},\mathbf{F}^{\text{B}})+ \sigma_b^2\mathbf{I}_{m_f}\right]^{-1} \mathbf{K}^{\text{B}}(\mathbf{F}^{\text{B}},\bar{\mathbf{f}}^{\text{B}}_*),
\end{aligned}
\end{equation} 
respectively.

\section{Closed Loop System for Behavioral Performance Induction }\label{Behavioral Performance Closed Loop System}
The bidirectional causality between the emotions and behavioral performance can be utilized to devise techniques that leverage available emotion elicitation techniques to adjust the quality of performance. In this section, we present a closed loop system that implements the behavioral performance inference presented in Section~\ref{Modeling of Behavioral Performance} to estimate the behavioral performance in real-time, and if necessary, introduce the required emotional stimulus to improve its quality. 

\subsection{Problem Formulation} \label{problem_statement_2}
Let the EEG-based features for a human operator, denoted by $\mathbf{e}_{k} \in \mathbb{R}^n$, be sampled at sampling time $t_k=k\Delta t, k=1,\dots,m+1$. Let the associated PAD states, denoted by $\mathbf{f}_k \in \mathbb{R}^3$ and the associated behavioral performance be denoted by $q_k \in \mathbb{R}$. Assume that the underlying process for the behavioral performance is described by,
\begin{equation}\label{perforamnce dynamics}
q_{k+1}=\mathfrak{f}^{\text{c}} (q_{k},\mathbf{u}_k), 
\end{equation}
where $\mathfrak{f}^{\text{c}}$ is an unknown map, and $\mathbf{u}_k \in \mathbb{R}^3$ is a vector of PAD influence of an external emotional stimulus.
Let the external emotional stimuli be a collection of audio trials represented by $\mathbf{M}^{\text{s}}=\{\mathbf{u}_{\text{s}_0},\dots,\mathbf{u}_{\text{s}_w}\}$, where $\mathbf{u}_{\text{s}_i} \in \mathbb{R}^3$, is the vector of PAD rating of the $i^{\text{th}}$ audio trial for $i\in \{1, \dots, w \}$, and $\mathbf{u}_{\text{s}_0} =\mathbf{0}$  refers to the trivial no audio trial. 
Let the targeted level of behavioral performance be $q_r$. Whenever $\mathbf{P}\{q_k \geqslant q_r\} < \beta_r$ for a given threshold $\beta_r \in (0,1)$, our objective is to deliver an audio stimulus ${\mu}_k$ such that $\mathbf{P}\{(q_{k+1}|q_k,\mu_k ) \ge q_r\}$ is maximized, i.e., identifying the following audio stimulus,
\begin{equation}\label{audio_stimulus}
    \mathbf{u}_k= k_r \hat{\mathbf{u}}_{s},
\end{equation}
where
\begin{eqnarray}
 k_r &=
    \begin{cases} \label{prob_cond}
      1 , & \mathbf{P}\{q_k \geqslant  q_r\} < \beta_r\\
      0 , & \text{otherwise}
    \end{cases},\\
     \nonumber \\
  \hat{\mathbf{u}}_{s}&= \underset{{{\mathbf{u}}_{\text{s}}\in \mathbf{M}^{\text{s}}}}{\text{arg max}}~~\mathbf{P}\{ (q_{k+1}|q_k,\mathbf{u}_k)\ge q_r\}.
\end{eqnarray}

\subsection{Behavioral Performance Closed Loop System}\label{CL_system}
Accomplishing the objective in Section~\ref{problem_statement_2} involves two critical ingredients: identifying an error function that describe the deviation of the current behavioral performance from the required level, and defining a function that maps the error into a control action which minimizes the deviation in future.
\\Let $\beta_k$ be the probability $\mathbf{P} \{ q_k \geqslant q_r \}$ at sampling time $t_k$. Then, an error function $\varepsilon_k$ can be defined by,
\begin{equation}
\begin{aligned} \label{error_func}
       \varepsilon_k &= & \beta_k-\beta_r &= & \mathbf{P} \{ q_k \geqslant q_r \}-\beta_r.
\end{aligned}
\end{equation}
An upper bound $\bar{\epsilon}_k$ for the error function in~\eqref{error_func} can be obtained using Markov inequality as follows,
\begin{equation}
\begin{aligned} \label{upper_error_func}
       \varepsilon_k &= & \mathbf{P} \{ q_k \geqslant q_r \}-\beta_r &\leqslant& \frac{\mathbf{E}\left[q_k\right]}{q_r}-\beta_r &=& \bar{\varepsilon}_k,
\end{aligned}
\end{equation}
where $\bar{q}_k \triangleq \mathbf{E}\left[q_k\right]$ is the mean function defined in~\eqref{behavioral_perf}. Define the standardized behavioral performance $\bar{\beta}_k\triangleq \frac{\bar{q}_k}{q_r}$, which represents the ratio between the expected and required behavioral performance. Then, the upper bound for the error function in~\eqref{upper_error_func}, namely the standardized behavioral performance error, can be written as,
\begin{align} \label{upper_ratio_error_func}
       \bar{\varepsilon}_k= \bar{\beta}_k-\beta_r .
\end{align}
\begin{figure}
\centering
\includegraphics[width=3in] {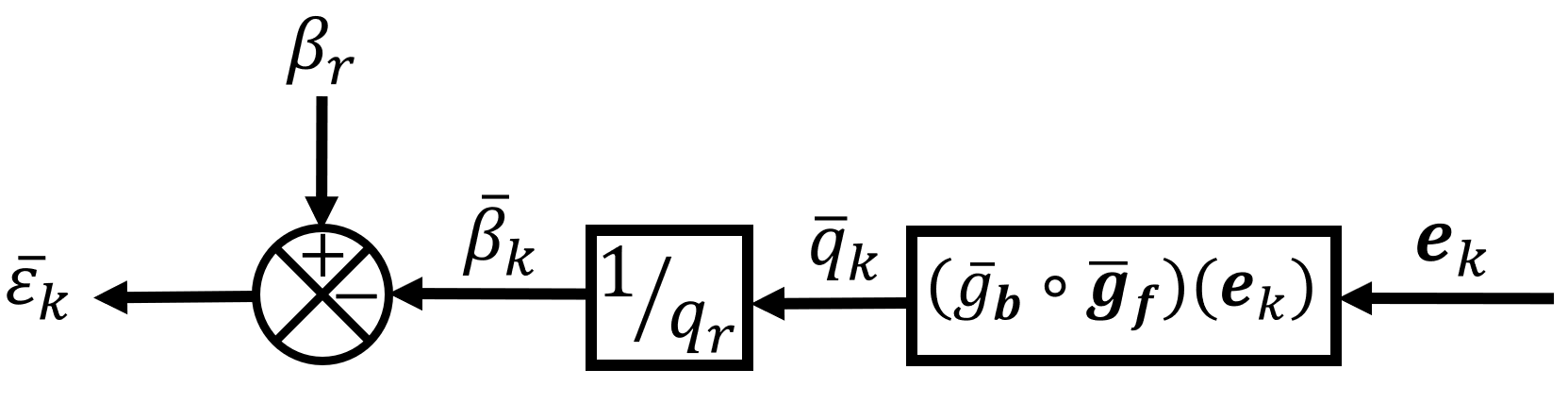}
\caption{Block diagram for the generation of the upper bound for the behavioral performance error function.}
\label{behavioral_perf_err}
\end{figure}
The generation of the upper bound for the error function $\bar{\varepsilon}_k$ is depicted by the block diagram in Figure~\ref{behavioral_perf_err}. The deterministic approximation of the error function defined by the equation~\eqref{upper_ratio_error_func} enables the utilization of numerous methods from control theory to design a controller that maps the generated error $\bar{\varepsilon}_k$ at time $t_k$ to the PAD state rating (value) $\hat{\mathbf{u}}_k$ for the audio stimulus to be delivered.

However, in the controller design techniques, it is essential for the fulfilment of control objectives a beforehand knowledge of precise mathematical model for the system of interest. Primarily, the performance of the designed controllers relies on level of resemblance between the spatiotemporal behavior predicted by the mathematical model in hand and that of the true system. Acquiring a satisfactorily accurate mathematical representation is a challenging task due to the complexity and uncertainty that characterize the behavior of the vast extent of real-world systems including~\eqref{perforamnce dynamics}.

\subsection{Fuzzy Logic Control}\label{fuzzy_system}
Fuzzy logic control is a heuristic method that implements human's approach of knowledge realization and decision making in the design of a nonlinear controller. At the core of fuzzy logic control is the concept of fuzzy sets which serve a mathematical framework to attain the symbolic and imprecise modeling that underscore the human capability in tolerating uncertainty in data. Fuzzy sets enable the representation of expert knowledge, which compensates the imperfect theoretical knowledge obtained by the existing practical experience. Thus, the ability of fuzzy logic controllers to process qualitative information and tolerate imprecision and uncertainty, makes their implementation favorable for systems with incomplete mathematical model or imprecise control action.

At the heart of the fuzzy systems is the utilization of fuzzy sets and fuzzy logic as they provide the mathematical framework for the representing numerical data qualitatively using linguistic variables and modeling the knowledge of experts using fuzzy rules. Fuzzy sets are structured to expand the conventional representation of the classical sets for physical quantities by adding a semantic evaluation to the numerical value using linguistic variables. Linguistic variables can be evaluated quantitatively using a numerical value and qualitatively using linguistic expression. A key benefit of linguistic variables is that it facilitates the representation of knowledge of expert using linguistic expression in the system while performing the basic functions of signal acquisition and generation, and data processing in term the numerical values. 
Fuzzy logic complements the rule of fuzzy sets by setting the framework for representing the knowledge of human expert and processing the qualitative information in real time. In particular, fuzzy logic defines the action for the union, intersection, and negation operators on the fuzzy sets, which is essential in modeling the knowledge of experts as a set of linguistic IF-THEN statements known as fuzzy rules, and in generating a control action by evaluating the fuzzy rules in real time.

\begin{figure}
\centering
\subfloat[]
{
{\includegraphics[width=2in] {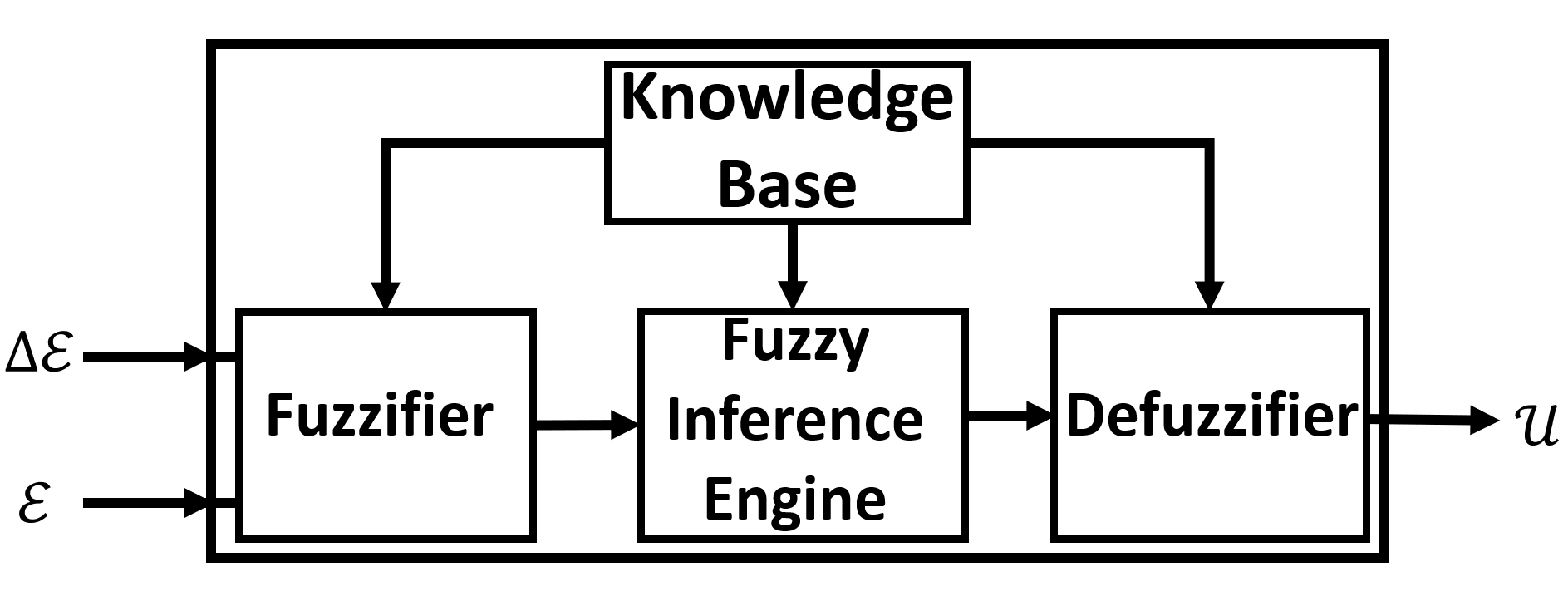}}
\label{flc_controller}
}
\hspace{0.3cm}
\subfloat[]
{
{\includegraphics[width=2in] {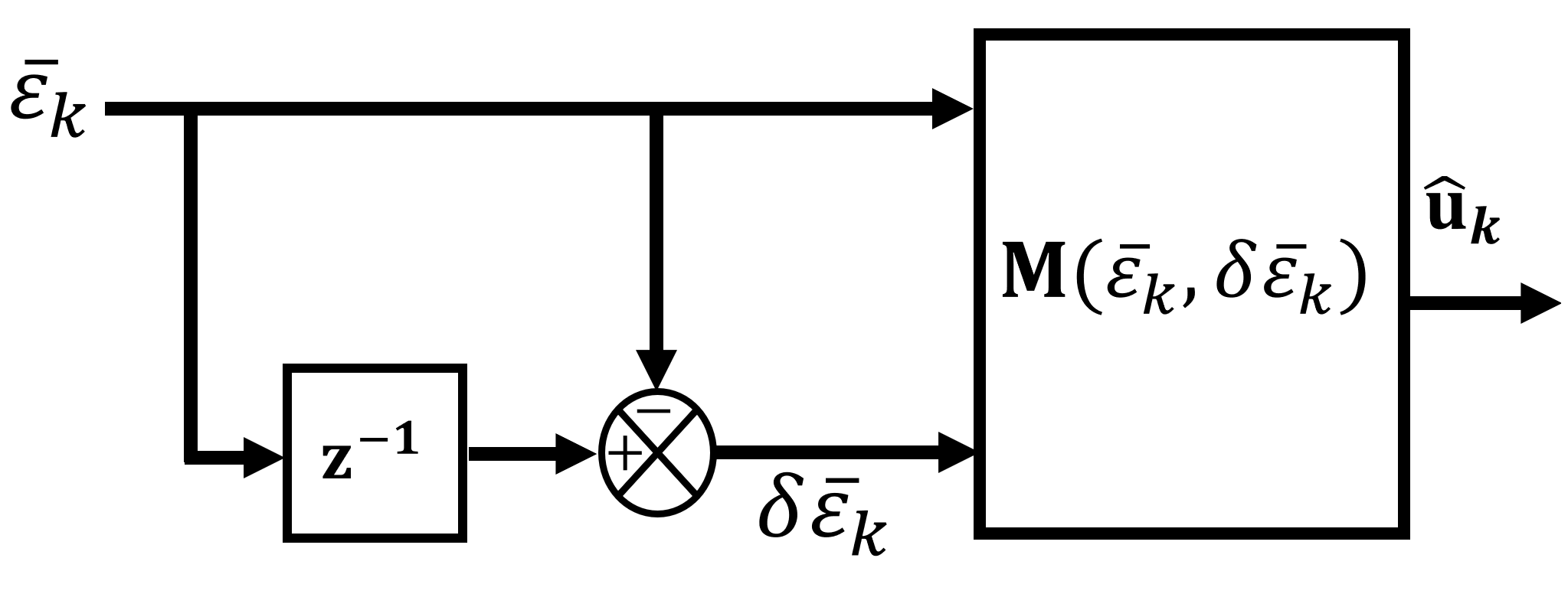}
\label{flc_behavioral_perf}}
}
\caption{(a)Structural architecture for a fuzzy logic controller, and (b) Block diagram for the behavioral performance fuzzy logic controller,}
\end{figure}

\subsubsection{Fuzzy Logic Controller}
Fuzzy logic controller~\cite{lee1990fuzzy} is knowledge-based system that encode the knowledge of experts as a nonlinear mapping of error $\mathcal{E}$ and change in error $\Delta\mathcal{E}$ into a control action $\mathcal{U}$, as it can be seen in Figure~\ref{flc_controller}. In general, four components constitute a fuzzy logic controller, namely,   
\begin{enumerate}
    \item Knowledge Base\\
    The knowledge base contains the knowledge of experts encoded as fuzzy rules which are a set of IF-THEN statements, and fuzzy sets for the error $\mathcal{E}$, change of error $\Delta\mathcal{E}$, and control action $\mathcal{U}$ (membership functions and associated linguistic values).
    \item Fuzzifier\\
    Given the fuzzy sets in the knowledge base, the fuzzifier converts the error and the change in error from their numerical states $\mathcal{E}$ and $\Delta \mathcal{E}$ into the corresponding fuzzy states $\tilde{\mathcal{E}}$ and $\Delta \tilde{\mathcal{E}}$, respectively.
    \item Fuzzy Inference Engine\\
    Fuzzy inference engine maps the fuzzy states $\tilde{\mathcal{E}}$ and $\Delta \tilde{\mathcal{E}}$ into a fuzzy state of control action $\tilde{\mathcal{U}}$. The mapping consists of evaluating the fuzzy rules according to fuzzy states of the error and change of error in the knowledge base, and making inference about the fuzzy state of the control action.
    \item Defuzzifier\\
    Finally, the fuzzy sets for the control action in the knowledge base are used to convert the fuzzy state of control action $\tilde{\mathcal{U}}$ into the corresponding numerical state $\mathcal{U}$ using the defuzzifier.  
\end{enumerate}

\subsection{Fuzzy Logic Control System for Behavioral Performance}
\begin{figure}
\centering
\subfloat[]
{
\includegraphics[width=1.5in] {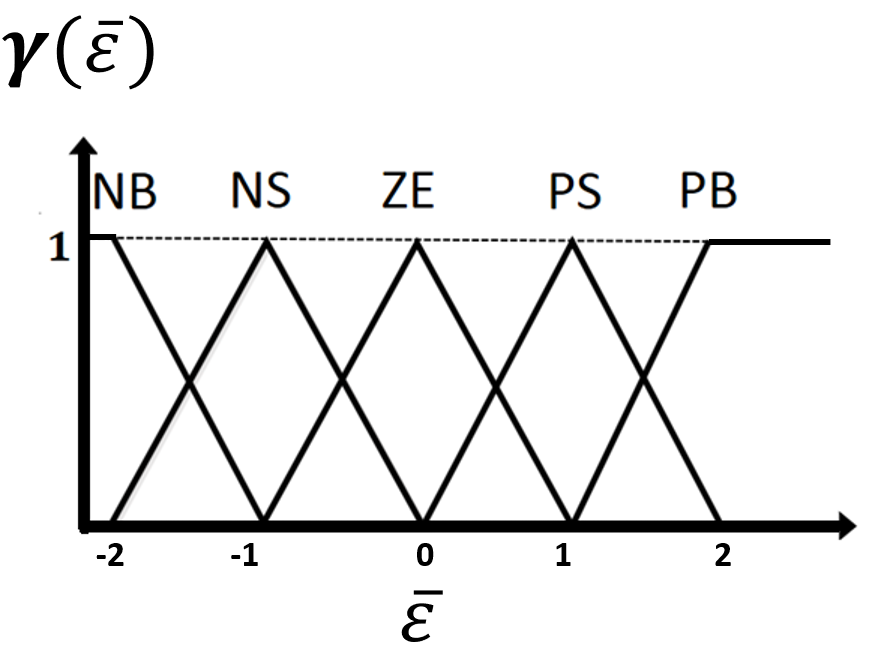}
\label{e-membership}
}
\subfloat[]
{
\includegraphics[width=1.5in] {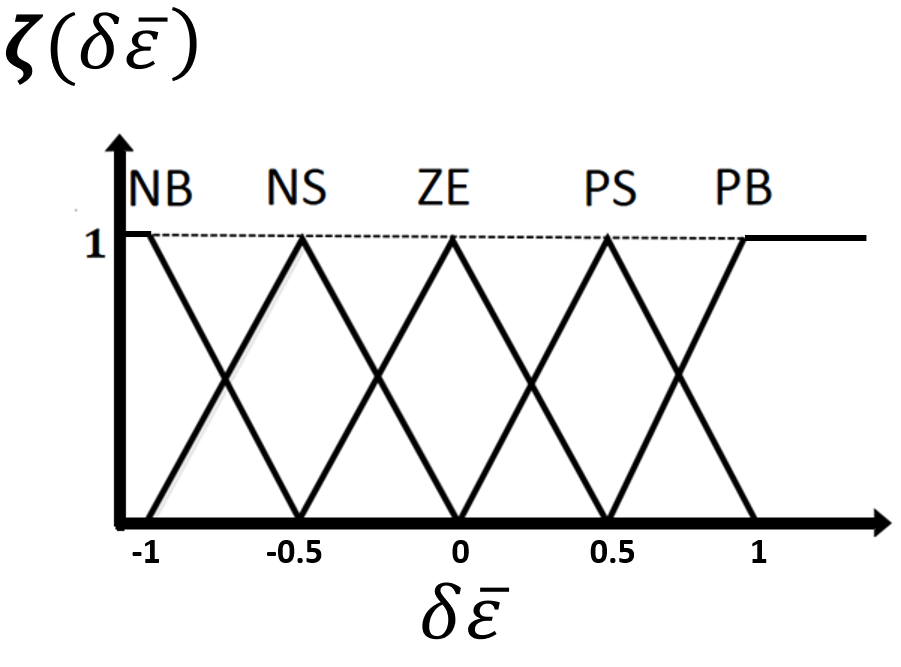}
\label{deltae-membership}
}
\\
\subfloat[]
{
\includegraphics[width=1.5in] {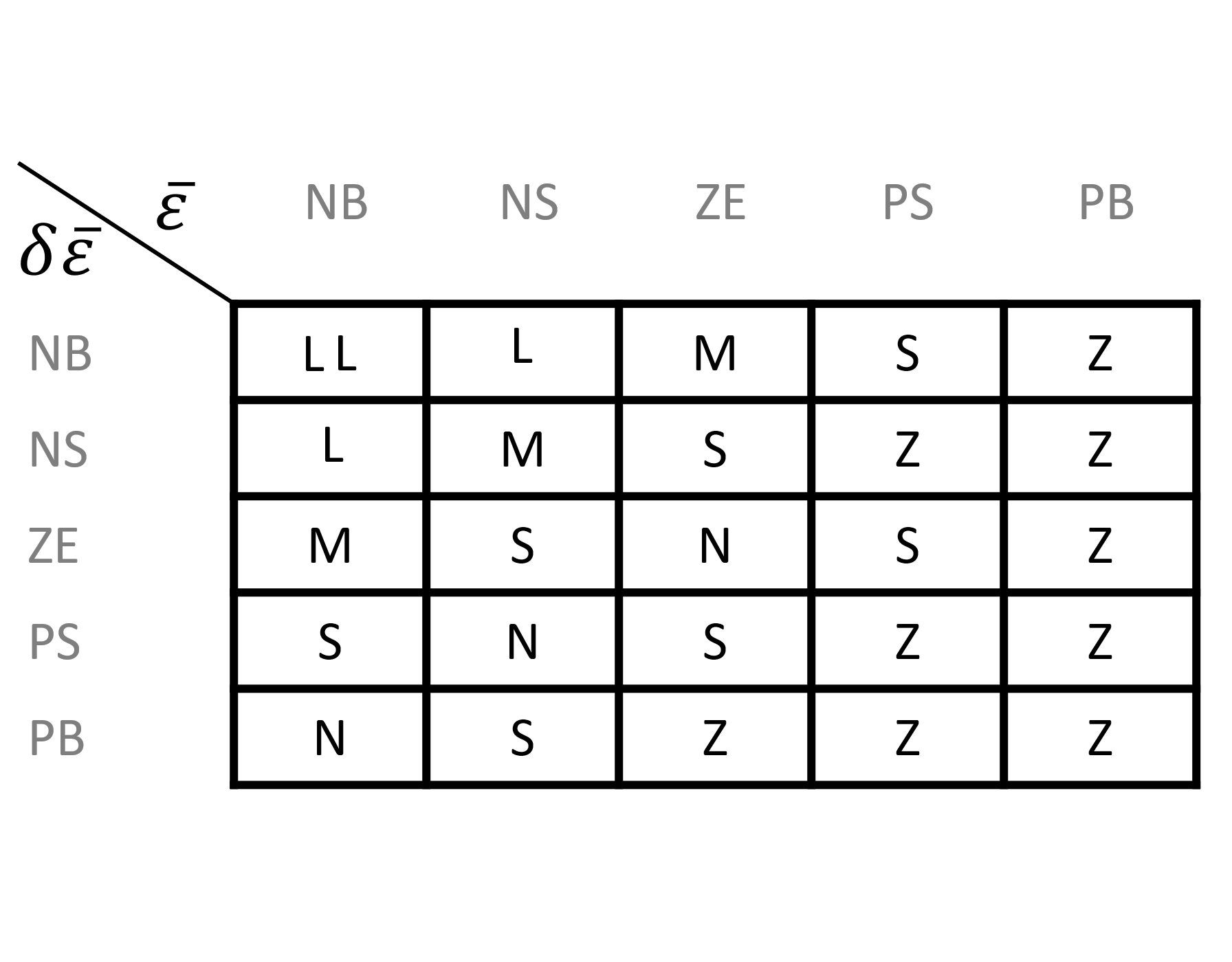}
\label{stim-membership}
}
\subfloat[]
{
\includegraphics[width=1.75in] {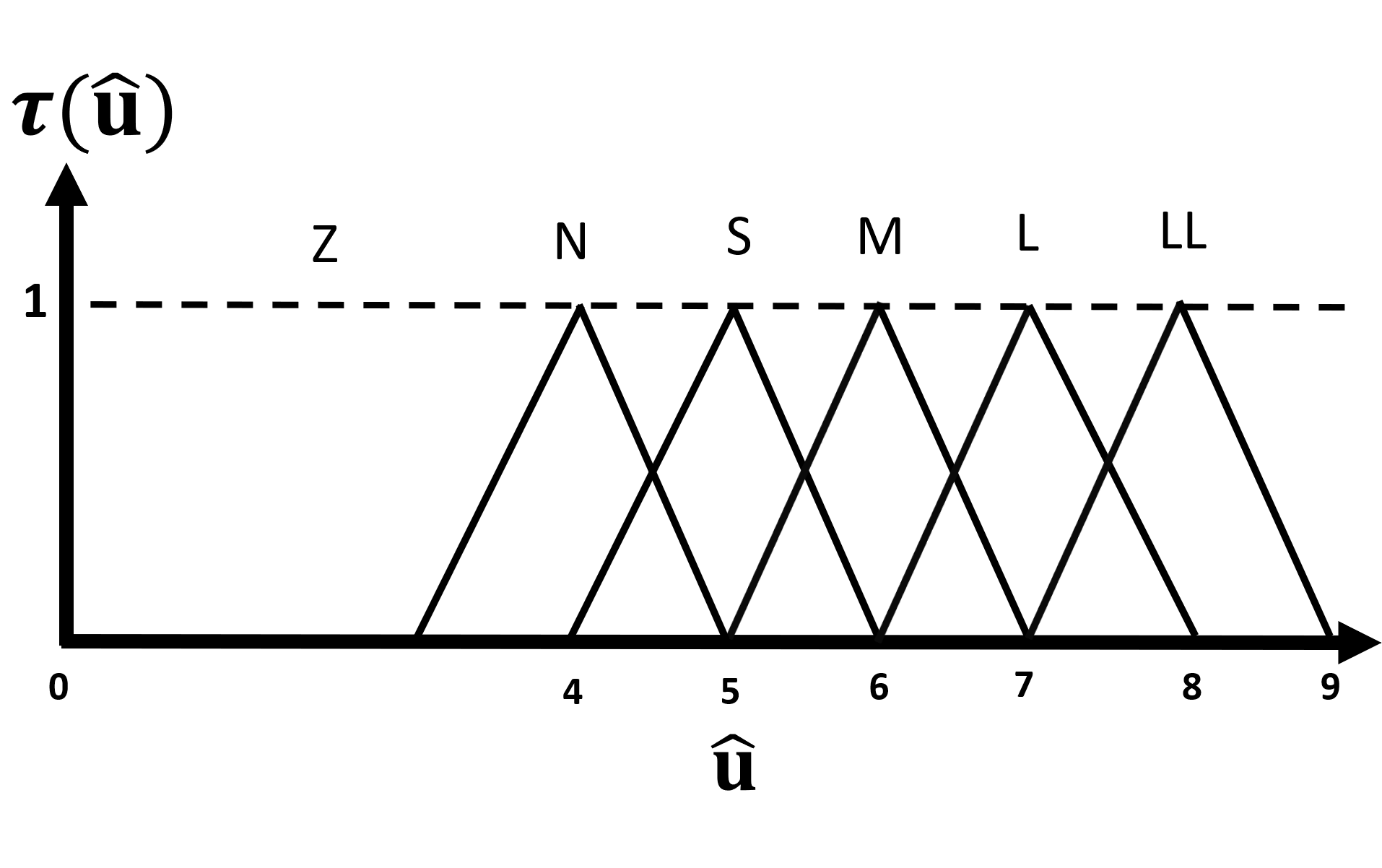}
\label{fuzzy_rule}
}
\caption{(a) Membership functions for the error, (b) Membership functions for the change in error, (c) Look up table for the fuzzy rules, (d) Membership function for the PAD values for the audio stimulus. }
\end{figure}

The selection of fuzzy logic control for the design of behavioral performance closed loop system is motivated by the following two reasons: first, the intra-individual variability in performance, which is associated with circumstantial factors such task difficulty, interest, and effort~\cite{fisher2004within}, makes the fuzzy sets suitable method for representing the quality of behavioral performance; second, the wealth of findings from the psychological studies in behavioral performance describing the qualitative impact for emotions on the behavioral performance can considered as the knowledge of experts to obtain a set of linguistic IF-THEN rules. Hence, the goal is to design a fuzzy logic controller $\mathbf{M}(\bar{\varepsilon}_k,\delta\bar{\varepsilon}_k)$, depicted in Figure~\ref{flc_behavioral_perf}, that maps the standardized error $\bar{\varepsilon}_k$ and change in standardized error $\delta\bar{\varepsilon}_k$ in behavioral performance into a PAD ratings of an audio stimulus $\hat{\mathbf{u}}_k$ that is defined in~\eqref{audio_stimulus}.   
\\ A substantial body of research has indicated that the task performance may be facilitated by positive emotions~\cite{fisher2004within}, which are associated with each element in PAD state vector satisfying the condition $\mathbf{f}_i > 4.5$, for $i \in \{1,2,3\}$. This finding can be stated as an implication by the following 
\\Let $q_k''$ and $q_k'$ be the behavioral performance at sampling time $t_k$ associated with the PAD states $\mathbf{f}_k''$ and $\mathbf{f}_k'$, respectively. Let $q_r$ be the targeted level of behavioral performance, and $\mathbf{f}_r=\begin{bsmallmatrix} 4.5 & 4.5 & 4.5 \end{bsmallmatrix}^{\text{T}}$ be the associated neutral PAD states. Then,
\begin{eqnarray}\label{implication1}
   \mathbf{E}[q_k''] \geqslant \mathbf{E}[q_k'] \longrightarrow \mathbf{E}[\mathbf{f}_k'']\geqslant \mathbf{E}[\mathbf{f}_k']
   &\implies & \mathbf{E}[q_k'']-{q_r}\beta_r \geqslant \mathbf{E}[q_k']-{q_r}\beta_r \nonumber \\
   \longrightarrow \mathbf{E}[\mathbf{f}_k''] -\mathbf{f}_r\geqslant \mathbf{E}[\mathbf{f}_k'] -\mathbf{f}_r 
  &\implies & \frac{\mathbf{E}[q_k'']}{q_r}-\beta_r \geqslant \frac{\mathbf{E}[q_k']}{q_r}-\beta_r \nonumber \\
  \longrightarrow \mathbf{E}[\mathbf{f}_k'']-\mathbf{f}_r \geqslant \mathbf{E}[\mathbf{f}_k'] -\mathbf{f}_r,
\end{eqnarray}
 where $\longrightarrow$ is the implication operator. Define $\varepsilon_k'' \triangleq \frac{\mathbf{E}[q_k'']}{q_r}-\beta_r$ and $\varepsilon_k' \triangleq \frac{\mathbf{E}[q_k']}{q_r}-\beta_r$ as the differential behavioral performances with respect to the targeted behavioral performance $q_r$, associated with differential PAD states $\mathbf{u}_k''\triangleq\mathbf{E}[\mathbf{f}_k'']-\mathbf{f}_r$ and $\mathbf{u}_k' \triangleq \mathbf{E}[\mathbf{f}_k']-\mathbf{f}_r$ with respect to the neutral PAD states $\mathbf{f}_r$. Then, the implication in~\eqref{implication1} can be written as
 \begin{eqnarray}\label{implication2}
    \varepsilon_k'' \geqslant \varepsilon_k' \longrightarrow \mathbf{u}_k''   \geqslant \mathbf{u}_k'. 
 \end{eqnarray}
As we described earlier, fuzzy sets provide the mathematical framework to represent the variation in behavioral performance and to model the implication in~\eqref{implication2} as a set of linguistic rules. Thus, we consider the fuzzy sets $\boldsymbol{\gamma}= \{\left(\bar{\varepsilon},\mu\left(\bar{\varepsilon}\right)\right) |\bar{\varepsilon} \in \mathcal{U}_{\gamma}\}$ and $\boldsymbol{\zeta}= \{\left(\delta\bar{\varepsilon},\pi\left(\delta\bar{\varepsilon}\right)\right) |\delta\bar{\varepsilon} \in \mathcal{U}_{\zeta}\}$, defined on the universes $\mathcal{U}_{\gamma}$ and $\mathcal{U}_{\zeta}$ of $\bar{\varepsilon}$ and $\delta\bar{\varepsilon}$, respectively. Furthermore, we define the fuzzy partition $\{\gamma_1, \cdots, \gamma_{5}\}$ of the fuzzy set $\boldsymbol{\gamma}$ and the fuzzy partition $\{\zeta_1, \cdots, \zeta_{5}\}$ of the fuzzy set $\boldsymbol{\zeta}$ characterized by the sets of pairs of membership function and linguistic value $(\boldsymbol{\mu},\boldsymbol{\rho})$ and $(\boldsymbol{\pi},\boldsymbol{\nu})$, respectively, which are defined as
\begin{eqnarray}
    (\boldsymbol{\rho},\boldsymbol{\mu})&=&\{(NB,\mu_1),(NS,\mu_2),(ZE,\mu_3),(PB,\mu_4),(PB,\mu_5)\}, \nonumber \\
    (\boldsymbol{\nu},\boldsymbol{\pi})&=&\{(NB,\pi_1),(NS,\pi_2),(ZE,\pi_3),(PB,\pi_4),(PB,\pi_5)\}, \nonumber
\end{eqnarray}
where $NB$, $NS$, $ZE$, $PS$, and $PB$, refer to Negative Big, Negative Small, Zero, Positive Small, and Positive Big, respectively. The membership functions along with their linguistic values for the error and change in error are shown in Figure~\ref{e-membership} and Figure~\ref{deltae-membership}, respectively. 
For the control action $\mathbf{u}$ defined as the PAD rating of the audio stimulus, we consider the fuzzy set $\boldsymbol{\lambda}= \{\left(\mathbf{u},\mu\left(\mathbf{u}\right)\right) |\mathbf{u} \in \mathcal{U}_{\lambda}\}$ defined on the universe $\mathcal{U}_{\lambda}$ of the control action, which is covered by the fuzzy partition $\{\lambda_1, \cdots, \lambda_{5}\}$. The later is characterized by the following set, 
\begin{equation}
    (\boldsymbol{\upsilon},\boldsymbol{\tau})=\{(Z,\tau_1),(N,\tau_2),(S,\tau_3),(M,\tau_4),(L,\tau_5),(LL,\tau_6)\}, \nonumber 
\end{equation}
where $Z$, $N$, $S$, $M$, $L$, and $LL$ refer to Zero, Neutral, Small, Large, Large Large, respectively. In accordance with the implication in~\eqref{implication2}, a map from the fuzzy state of error and change in error to the fuzzy state of the control action $\tilde{\mathbf{u}}$ is defined as a set of 25 linguistic rules written as,
\begin{align}\label{fuzzy_rules_set_beh}
    \mathcal{R}^{(i,j)}_l= \Bigg\{ \text{IF} \left(\bar{\varepsilon}  ~~\text{is}~~ \nu_{i}\right) ~\text{and}~  \left(\delta\bar{\varepsilon} ~~\text{is}~~ \rho_{j}\right), ~\text{THEN} ~ \left(\tilde{\mathbf{u}} ~\text{is}~ \boldsymbol{\upsilon}_l^{(i,j)}\right) \Bigg\}^{5}_{i,j=1}, 
\end{align}
where $\boldsymbol{\upsilon}_l^{(i,j)}$ is the $(i,j)$ element in the look up table shown in Figure~\ref{fuzzy_rule} and associated with the $l^{\text{th}}$ linguistic value in $\boldsymbol{\upsilon}$ . The fuzzy subset $\lambda_{1}$ is assigned a zero membership function $\tau_1(\mathbf{u})=\mathbf{0}$, to prevent any control action when the behavioral performance is above the required level. 

The fuzzy sets and rules presented above constitute the knowledge base component which is required for the fuzzy logic controller $\mathbf{M}(\bar{\varepsilon}_k,\delta\bar{\varepsilon}_k)$ to be able to execute the following steps at each sampling time $t_k$ 
\begin{enumerate}
    \item Fuzzification\\
    Given the fuzzy subsets $\{\gamma_1, \cdots, \gamma_{5}\}$ and $\{\zeta_1, \cdots, \zeta_{5}\}$, the error and change in error are converted from their numerical states ${\varepsilon}_k$ and $\delta\bar{\varepsilon}_k$ to the corresponding fuzzy states $\tilde{{\varepsilon}}_k$ and $\delta{\tilde{{\varepsilon}}}_k$ defined as
    \begin{eqnarray}
       \tilde{{\varepsilon}}_k=&(\boldsymbol{\nu}_k, \boldsymbol{\mu}(\bar{\varepsilon}_k))=&
        \begin{bsmallmatrix} \left(NB,\mu_1(\bar{\varepsilon}_k)\right)& 
        \left(NS,\mu_2(\bar{\varepsilon}_k)\right)& 
        \left(ZE,\mu_3(\bar{\varepsilon}_k)\right) &
        \left(PS,\mu_4(\bar{\varepsilon}_k)\right)&
        \left(PB,\mu_5(\bar{\varepsilon}_k)\right) 
       \end{bsmallmatrix}, \nonumber \\
       \delta\tilde{\varepsilon}_k=&(\boldsymbol{\upsilon}_k, \boldsymbol{\pi}(\delta\bar{\varepsilon}_k))=&
        \begin{bsmallmatrix} \left(NB,\pi_1(\delta\bar{\varepsilon}_k)\right) & \left(NS,\pi_2(\delta\bar{\varepsilon}_k)\right)& \left(ZE,\pi_3(\delta\bar{\varepsilon}_k)\right) & \left(PS,\pi_4(\delta\bar{\varepsilon}_k)\right)& \left(PB,\pi_5(\delta\bar{\varepsilon}_k)\right) 
       \end{bsmallmatrix}\nonumber ,
    \end{eqnarray}
    \item{Fuzzy Inference}\\
     Given the set of rules in~\eqref{fuzzy_rules_set_beh}, and the fuzzy states $\tilde{{\varepsilon}}_k$ and $\delta{\tilde{\varepsilon}}_k$, an inference for the fuzzy state $\tilde{\mathbf{u}}_k$ is obtained according to the following
     \begin{eqnarray} \label{perf_fuzzy_rules}
     \tilde{\mathbf{u}}_k =&(\boldsymbol{\upsilon}_k, \boldsymbol{\tau}(\hat{\mathbf{u}}_k))=&
        \begin{bsmallmatrix} \left(Z,\tau_1(\hat{\mathbf{u}}_k)\right) & \left(N,\tau_2(\hat{\mathbf{u}}_k)\right)& \left(S,\tau_3(\hat{\mathbf{u}}_k)\right) & \left(M,\tau_4(\hat{\mathbf{u}}_k)\right)& \left(L,\tau_5(\hat{\mathbf{u}}_k)\right) & \left(LL,\tau_6(\hat{\mathbf{u}}_k)\right)  
        \end{bsmallmatrix}\nonumber,
     \end{eqnarray}
     where for $l\in\{1,\cdots,6\}$,
    \begin{equation}\label{memebership_values_u}
    \begin{aligned}
       {\tau}_l(\hat{\mathbf{u}}_k)=&\overset{5}{\underset{i,j=1}{\bigvee}}{\tau}^{(i,j)}_l(\hat{\mathbf{u}}_k),& 
       {\tau}^{(i,j)}_l(\hat{\mathbf{u}}_k)=&\mu_{i}\left(\bar{\varepsilon}_k\right) \bigwedge \tau_{j}\left(\delta\bar{\varepsilon}_k\right),  
     \end{aligned} 
     \end{equation}
     and the triangular conorm operator $ \bigvee :[0,1]\times[0,1]\rightarrow[0,1]$, and the triangular norm operator $ \bigwedge :[0,1] \times [0,1] \rightarrow [0,1]$, respectively, defined as follows
     \begin{eqnarray}
        \bigvee \left(\mathfrak{f}(a) , \mathfrak{f}(b) \right)=\text{max} \left(\mathfrak{f}(a) , \mathfrak{f}(b) \right), ~  \bigwedge \left(\mathfrak{f}(a) , \mathfrak{f}(b)\right)=\text{min} \left(\mathfrak{f}(a) , \mathfrak{f}(b) \right), \nonumber
     \end{eqnarray}
     \item Defuzzification\\
     Given the fuzzy subsets $\{\lambda_1,\cdots,\lambda_5\}$, the fuzzy state $\tilde{\mathbf{u}}_k$ calculated from~\eqref{perf_fuzzy_rules}, is converted to the numerical state $\hat{\mathbf{u}}_k$ using the following formula
    \begin{align} \label{coa}
        \hat{\mathbf{u}}_k= \frac{\sum\limits_{l=1}^{5}\tau_l(\hat{\mathbf{u}}_k) \mathbf{u}^{\text{c}}_{l}}{\sum\limits_{i=l}^{5} \tau_l(\hat{\mathbf{u}}_k)}, \nonumber
   \end{align}
   where $\tau_l(\hat{\mathbf{u}}_k)$ is calculated from ~\eqref{memebership_values_u} for $l\in\{1,\cdots,6\}$, and $\mathbf{u}^{\text{c}}_{l}= \underset{{\mathbf{u}}^{\text{s}}}{\text{arg}}~ \text{max}~ \tau_l(\hat{\mathbf{u}})$ achieves the maxima of the membership function $\hat{\mathbf{u}}_k$, for $l\in\{1,\cdots,6\}$.
\end{enumerate}

Let $\mathbf{M}^{\text{s}}=\{\mathbf{u}_1^{\text{s}}, \dots,\mathbf{u}_q^{\text{s}}\}$ be the set of PAD ratings for the available audio stimuli where $\mathbf{u}_i^{\text{s}} \in \mathbb{R}^3$, for $i\in \{1,\dots,q\}$. Then, the delivered audio stimulus $\tilde{{\mathbf{u}}}_k$ is defined by,
\begin{equation}
\begin{aligned}\label{audio_stim_selec} 
           \tilde{{\mathbf{u}}}_k&= \underset{{\mathbf{u}}^{\text{s}}}{\text{arg}}~ \text{min}~ \left\| \mathbf{u}^{\text{s}}- \hat{\mathbf{u}}_k \right\|_2, &\mathbf{\mathbf{u}}^{\text{s}}\in \mathbf{M}^{\text{s}}, 
\end{aligned}
\end{equation}
which approximates the required stimulus in equation~\eqref{audio_stimulus}.

\section{Experimental Setup}\label{experimental setup}
In this section, the three conducted experiments are explored. The purpose of the first two experiments, namely, emotion elicitation and behavioral performance induction, are to collect the necessary data to infer the predictive distribution of behavioral performance, as discussed in Section~\ref{Modeling of Behavioral Performance}. In the third experiment, the inferred distribution from the first two experiments is used to implement the closed loop system for behavioral performance induction which is described in Section~\ref{Behavioral Performance Closed Loop System}.     

\subsection{Material}
Fifteen graduate and undergraduate students participated in the first two experiment, and three graduate students participated in the third experiment. Their ages range between 18 and 30 years (Mean$=25\pm5$ years). Each participant had orientation session in which the methods  and goals of each experiment were explained in details. None of the participants reported any physiological/psychological disorder history, or  alcohol/drug abuse problem, which may affect their EEG signals. All participants signed the informed consent before the experiments which has the approval of the Institutional Review Board in the Human Research Protection Program at Michigan State University. 
\newline EPOC+ headset (EMOTIV, USA) was used to collect EEG signals from 14 channels during the three experiments. The mobility, ease of use, and wireless connectivity, make EPOC+ headset suitable for on-field robotic applications. Channel locations are based on the international 10-20 electrode location system, which are: AF3, F7, F3, FC5, T7, P7, O1, O2, P8, T8, FC6, F4, F8, AF4, with CMS/DRL references in the P3/P4 locations. The sampling rate is 128 Hz with resolution of 16 bits per channel. RoboWorks animation and modeling software was used to create virtual reality environment in which participants were asked to manipulate a robotic arm to perform predefined tasks .

\subsection{Emotion Elicitation Experiment}\label{emotion_elicitation}
\begin{figure*}[!t]
\centering 
\subfloat[ ]
{
\label{emotion_elicit}
\includegraphics[width=2.5in]{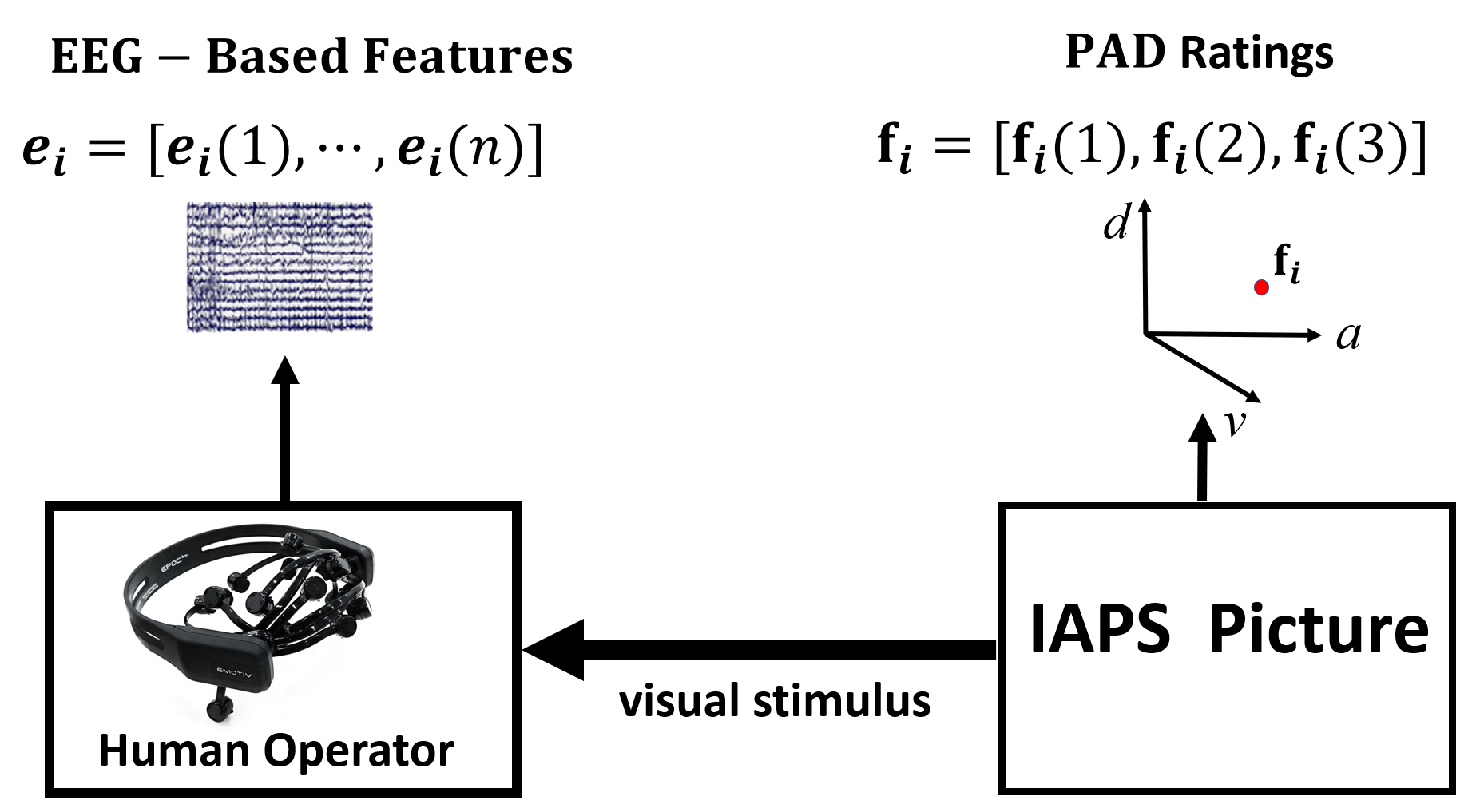}
}
\hspace{5em}
\subfloat[ ]
{
\label{performance_emotion_elicit}
\includegraphics[width=2.5in]{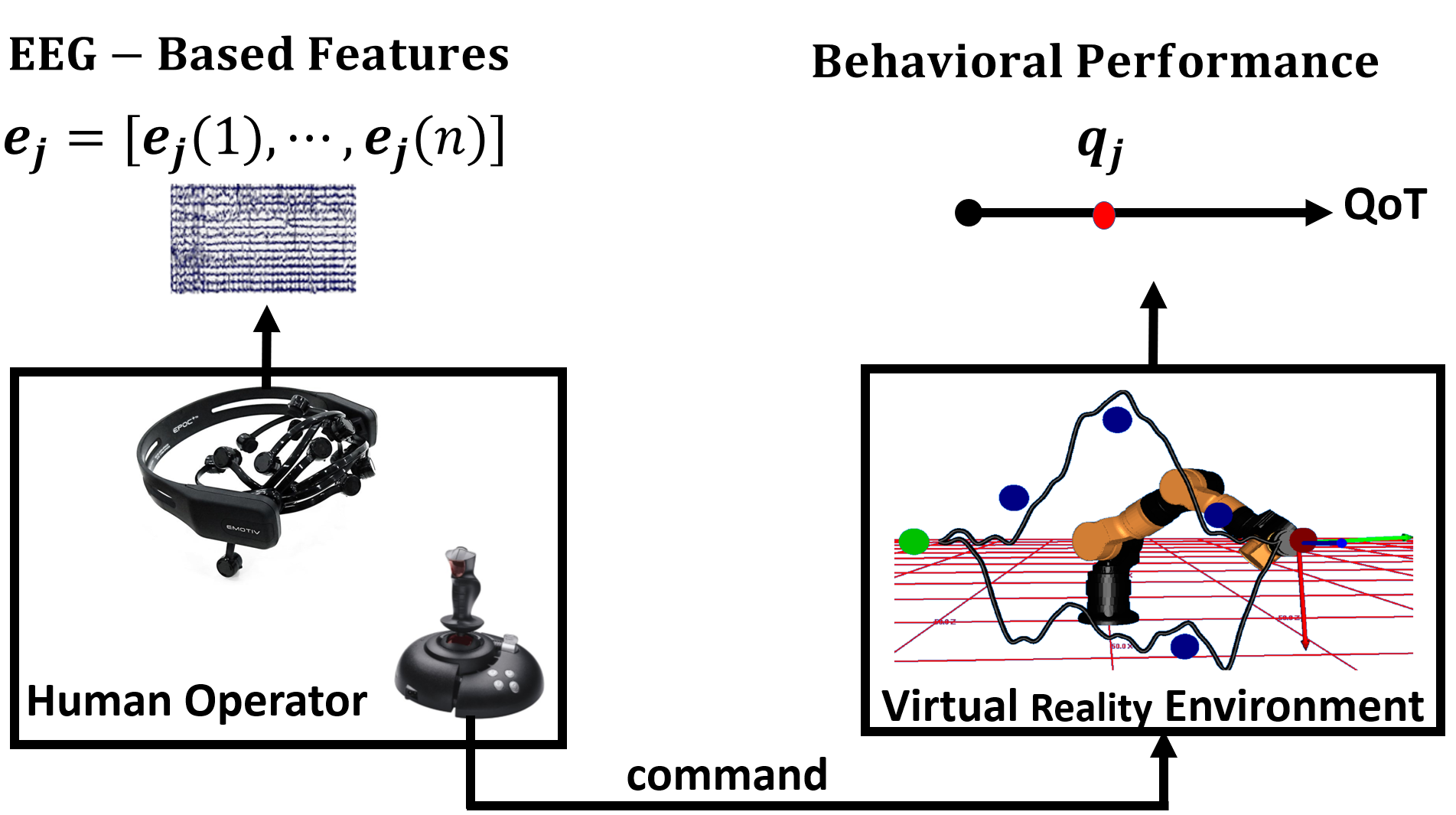}
}
\caption{(a)Data collection in the Emotion elicitation experiment during the exposure to the $i^{th}$ visual stimuli, for the EEG-based features $\mathbf{e}_i$ and the PAD ratings $\mathbf{f}_i$, and (b) data collection in the behavioral performance induction experiment during the $j^{th}$ for the EEG-based features $\mathbf{e}_j$ and the QoT value ${q}_j$.}
\centering
\end{figure*}

The objective of this experiment is to collect the datasets $\{\mathbf{e}^{\text{E}}_i,\mathbf{f}^{\text{E}}_i\}_{i=1}^{m_f}$ of EEG-based features vector $\mathbf{e}^{\text{E}}_i \in \mathbf{R}^n$ and the PAD states $\mathbf{f}^{\text{E}}_i \in \mathbb{R}^3$. Each participant had 3 different sessions during 3 separate days. The visual stimuli were selected from the International Affective Picture Systems (IAPS)~\cite{lang1997international}, which is a database which was established by the Center for Study of Emotion and Attention (CSEA) at University of Florida to provide PAD states ratings for a large set of emotionally evocative visual stimuli based on the dimensional model of emotions presented by Russell~\cite{russell1977evidence}.  Each rating is within the range of 1 to 9. The value 1 refers to completely unhappy, completely unaroused, and extremely controlled for valence, arousal and dominance, respectively. On the other end of the scale, the value 9 refers to completely happy, completely aroused, and extremely in control for valence, arousal and dominance respectively. Three sets of pictures with total of 183 pictures were selected. The three sets are mutually exclusive and each set is chosen to span the range from 1 to 9 for each emotional state.The emotion elicitation procedure presented in~\cite{koelstra2011deap} is adapted in this work.
In the beginning of each session, a black screen is displayed to the participant for 30 seconds to set EEG signal to the baseline level. After that, a white cross is displayed for 5 seconds to inform the participant that a visual stimulus would be shown. After the disappearance of the cross, a picture from IAPS database is displayed for 10 seconds while recording the EEG-based features $\mathbf{e}_i$ and the PAD ratings for the visual stimulus $\mathbf{f}_i$, as it is shown in Figure~\ref{emotion_elicit}. Upon the disappearance of the picture, another cross sign shows up for 5 second followed by another picture for 10 seconds, and so on. Visual C++ code was designed to synchronize the picture displaying and data recording.

\subsection{Behavioral Performance Induction Experiment} \label{performance_induction}
The objective of this experiment is to collect the dataset $D^{\text{B}}=\{\mathbf{e}_i,q_i\}_{i=1}^{m_b}$ of EEG-based features vector $\mathbf{e}_i \in \mathbf{R}^n$ and the behavioral performance measure $q_i \in \mathbb{R}$, required to make inference about the predictive distribution of behavioral performance discussed in Section~\ref{Modeling of Behavioral Performance}. In order to ensure that the datasets encode the impact of wide range of behavioral performance, the task difficulty and duration were tuned to induce fatigue-based degradation in behavioral performance. Also, the participants completed three subjective-fatigue and sleepiness scale questionnaire -- Stanford Sleepiness Scale (SSS), Samn-Perelli Scale (SPS), and Karolinska Sleepiness Scale (KSS) -- before and after the experiment to check the influence of fatigue on the observed performance.
Each participant was required to command an articulated robotic arm in a virtual reality environment to make the end effector follows a predefined set of trajectories $i\in \{1, \dots, 60\}$, and the EEG-based features $\mathbf{e}_i$ and the QoT value $q_i$ were recorded. The direction and speed of manipulator's end effector is controlled by the participant through USB-connected Microsoft Sidewinder Force Feedback 2 joystick. Visual C++ code was designed to deliver the commands from the joystick to manipulator in   the virtual reality environment, and to record EEG-based features and behavioral performance measures within a 10 seconds non-overlapping sliding window. 

\subsection{Behavioral Performance Regulation Experiment}
\begin{figure}[!b]
\centering
\includegraphics[width=5.0in]{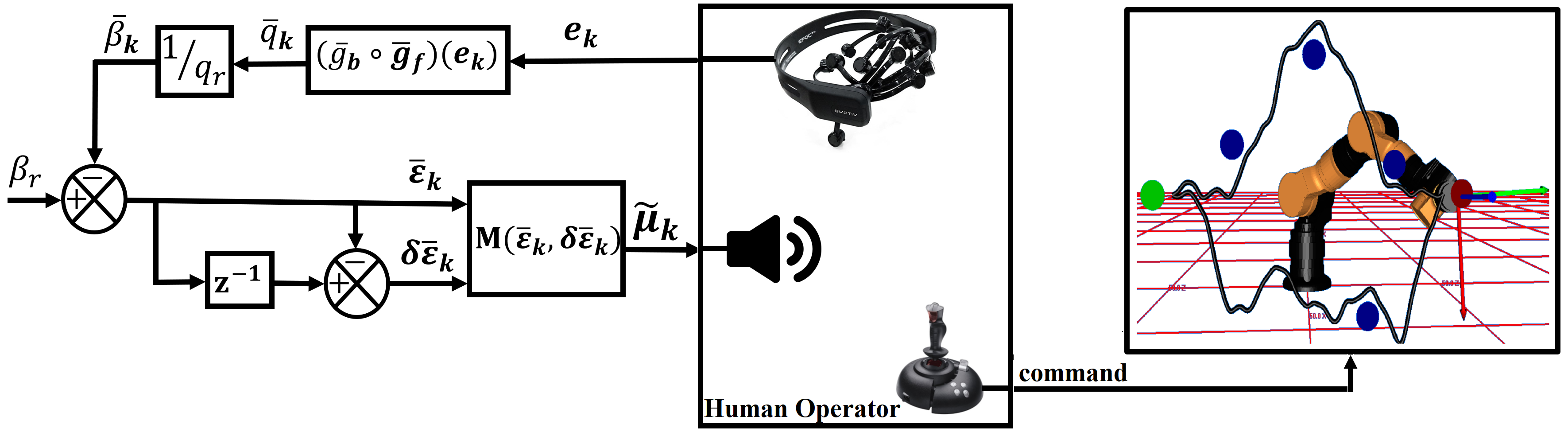}
\caption{Closed Loop System for Behavioral Performance Induction}
\label{CL_performance}
\end{figure}

In this experiment, the predictive distribution in~\eqref{behavioral_perf}, which is inferred from the datasets collected during the experiments presented in Sections~\ref{emotion_elicitation} and ~\ref{performance_induction}, is utilized to implement the closed loop system in Figure~\ref{CL_performance}.
    Each participant was required to command an articulated robotic arm in a virtual reality environment to make the end effector follow a predefined set of trajectories. The EEG-based features were recorded using the EPOC+ headset while the subject performed the experiment. At sampling time $t_k$, the recorded $\mathbf{e}_k$ is used to calculate $\bar{q}_k$ and $\bar{\beta}_k$ defined in~\eqref{upper_error_func}. Consequently, the error $\bar{\varepsilon}_k$ and the change in error $\delta\bar{\varepsilon}_k$ defined in~\eqref{upper_ratio_error_func} is applied to the fuzzy controller $M(\bar{\varepsilon}_k,\delta\bar{\varepsilon}_k)$ to obtain the control action $\hat{\mathbf{\mu}}_k$. A popular database, known as DEAP~\cite{koelstra2011deap}, is used to establish the set of audio stimuli $\mathbf{M}^{\text{s}}$, from which the audio stimulus $\tilde{\mu}_s$ defined in~\eqref{audio_stim_selec} is selected and delivered to the participant via pair of earphones. A C++ code is designed the collection of EEG signals from the EPOC+ headset, processing the EEG-based features to specify and play audio stimuli from $\mathbf{M}^{\text{s}}$, and communicating between the joystick and the manipulator in the virtual reality environment.

\section{Results and Discussion}\label{results and discussion}

\subsection{PAD states Inference} \label{PAD states Inference}
 There is a growing trend in Brain-Computer Interface (BCI) to use the complexity and chaotic measures for EEG signals as distinguishing features for medical and non-medical applications~\cite{lane2011physiological}. In this work, we build upon previous results~\cite{adeli2007wavelet, liu2013eeg} and use the Fractal Dimension (FD) values as features for the estimation of  PAD states and behavioral performance. We also investigate the spectral impact of the emotions and behavioral performance on the EEG signals by using two types of fractal features, which are the fractal dimension for EEG  and the fractal dimension for EEG bands to obtain two Gaussian Process (GP) regression models, namely, EEG-based GP model and EEG bands-based GP model. The two models are compared  in term of prediction accuracy.

\subsubsection{Fractal Features for EEG}\label{fractal feature}
The EEG data collected during the emotion elicitation and behavioral induction experiments were preprocessed in three steps as follows: first, Common Average Reference (CAR) filtering, second; Band Pass (BP) filtering $(1-60 \mathrm{Hz})$, and finally, Blind Source Separation (BSS) techniques were used to remove the eye artifacts~\cite{gomez2006automatic}. EEGLAB was used for implementing the preprocessing steps and generating the topographical views for the related brain activities~\cite{delorme2004eeglab}. 
For the decomposition of EEG signals into the rhythmic bands, a Discrete Wavelet Transform (DWT)-based filter was applied to leverage the scalability and time-frequency localization of wavelet transform  which make it more effective than Fourier transform-based filters in extracting features from non-stationary signals such as EEG signals~\cite{adeli2007wavelet,adeli2003analysis}. Hence, given that the sampling rate of the EPOC+ headset is $128\mathrm{Hz}$, a fourth order Daubechies wavelet is used as mother function to extract the detailed coefficients $D_1 (32-64~\mathrm{Hz})$, $D_2 (16-32~\mathrm{Hz})$, $D_3 (8-16~\mathrm{Hz})$, and $D_4 (4-8~\mathrm{Hz})$ which can be reconstructed-by using Inverse DWT– to approximate theta $(4-8~\mathrm{Hz})$, alpha $(8-16~\mathrm{Hz})$, beta $(16-32~\mathrm{Hz})$, and gamma $(32-64~\mathrm{Hz})$ bands. Finally, Higuchi fractal dimension method~\cite{higuchi1988approach} was used to generate two datasets of EEG-based features: EEG-based FD dataset which consists of the FD values of EEG signals at each electrode, and EEG bands-based FD dataset which consists of the FD values of EEG bands at each electrode.

\begin{figure}[!t]
\centering
\includegraphics[width=5.0in]{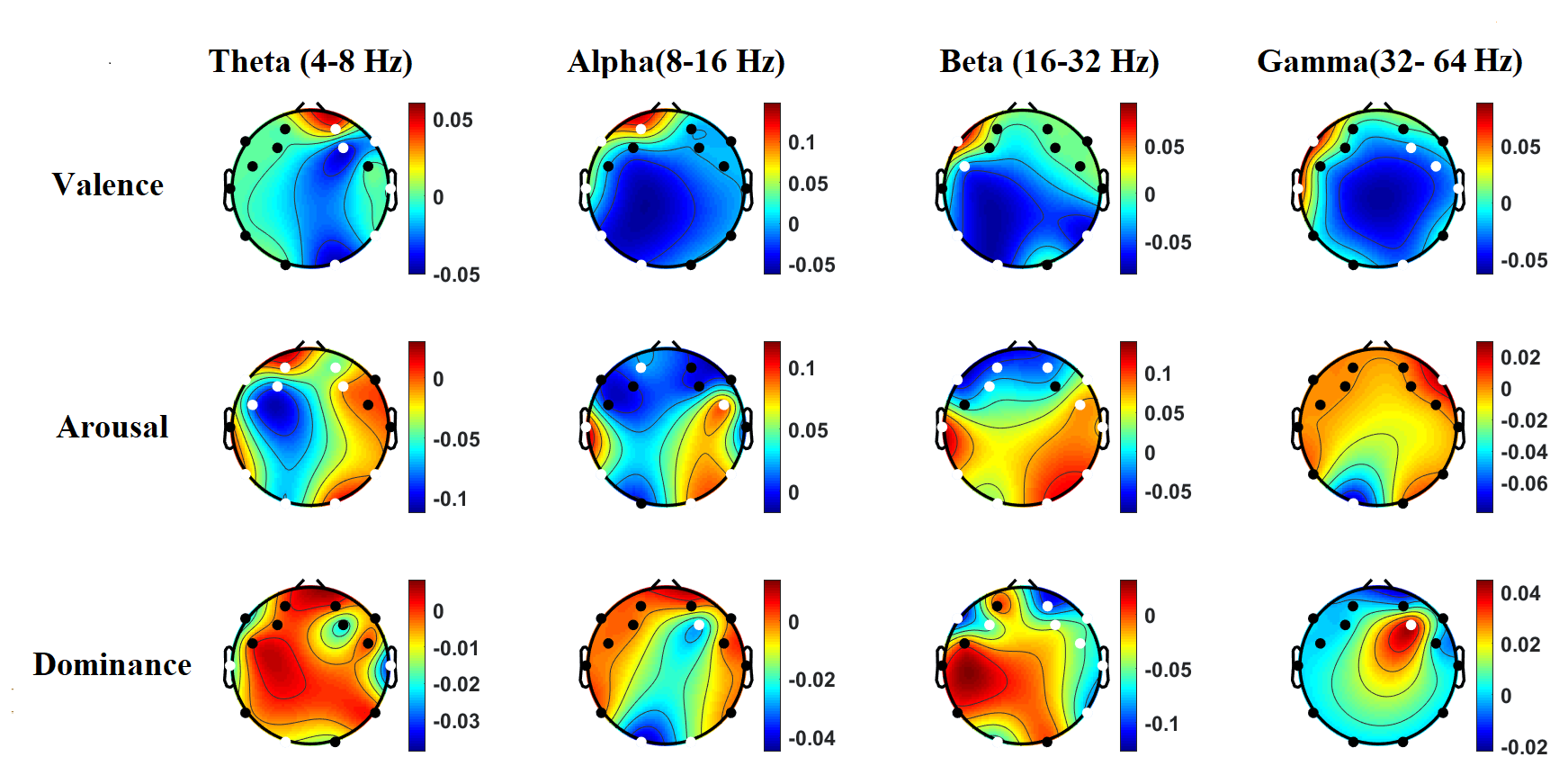}
\caption{Mean correlation over all participant between FD values for theta (4-8 Hz), alpha (8-16 Hz), beta (16-32 Hz), and gamma (32-64 Hz) and the three PAD states ratings from the IAPS. Electrode channels with significant correlation are depicted by white markers}
\label{5:correlation}
\end{figure}

In order to examine the statistical relationship between the fractal activity of EEG signals and the concurrent emotional state, we calculate the Spearman correlation $r$ for each participant between the FD values per channel and band and the PAD ratings from IAPS database. Furthermore, the statistical significance $p$ for the observed correlation $r$ was obtained by computing the p-value for the one-tailed tests (left and right). Finally, the correlation coefficients and significance for all participants were fused by taking the average of correlation coefficients $\mathrm{\bar{R}}$, and computing the combined ${\bar{p}}$ value via QFAST algorithm~\cite{bailey1998combining}. Detailed set of significant correlations (${\bar{p}}<0.05$) is presented in Table \ref{electrode sub-band correlation}, in which $\mathrm{\bar{R}}$ represents the mean correlation coefficients of all participants, $\mathrm{R^-}$ represents the most negative correlation value, and $\mathrm{R^+}$ represents the most positive correlation value. 
For valence, significant correlations were observed for all frequency bands in the frontal and temporal regions. According to these correlation values, the temporal and frontal regions of left hemisphere show more positive correlation with valence level, as opposed to the temporal and frontal regions of right hemisphere for  alpha, beta, and gamma bands. These observations are consistent with the results reported in~\cite{liu2013eeg}, which show that the left hemisphere has higher activity in term of FD values than the right hemisphere during high valence states. At the same time, the occipital and posterior regions for both hemisphere showed negative correlation with valence for all bands. For arousal, significant positive correlations are recorded for alpha and beta bands in the temporal, parietal and occipital regions which contains somatosensory, auditory, and visual cortices. 
An increasing processing rate of cognitive and sensory information is associated with irregularity in the activity of alpha and beta bands known as desynchronization ~\cite{pfurtscheller2001functional}. Hence, our results suggest that high level of arousal can be associated with a desynchronization in EEG activities. Finally, all significant correlations between dominance and FD values are negative in the frontal, central, parietal, temporal and occipital regions, which indicates that the feeling of dominance is accompanied with a an increase in the FD values. Figure \ref{5:correlation} shows the topographic distribution of the average correlations  $\mathrm{\bar{R}}$ over the EEG bands per channels for the three PAD states.
\begin{table}[t!]
\centering
 \renewcommand{\arraystretch}{1}
\caption{band per electrodes which have significant correlation between their FD values and IAPS ratings ($ ^*=\bar{p}<0.01, ^{**}=\bar{p}<0.001 $), $\mathrm{\bar{R}}$ is the mean correlation, $\mathrm{R^+}$ is the most positive correlation, $\mathrm{R^-}$ is the most negative correlation}
\label{electrode sub-band correlation}
\setlength{\arrayrulewidth}{0.1mm}
\setlength{\tabcolsep}{3pt}
\renewcommand{\arraystretch}{1.1}
\resizebox{\textwidth}{!}
{
\begin{tabular}{c c c c c c c c c c c c c c c c c }
\hline
\hline
\multicolumn{1}{c}{emotion} & \multicolumn{4}{c}{theta}  & \multicolumn{4}{c}{alpha} & \multicolumn{4}{c}{beta}  & \multicolumn{4}{c}{gamma} \\
 & Elec. & $\mathrm{R^-}$ & $\mathrm{\bar{R}}$ & $\mathrm{R^+}$ & Elec. & $\mathrm{R^-}$ & $\mathrm{\bar{R}}$ & $\mathrm{R^+}$ & Elec. & $\mathrm{R^-}$ & $\mathrm{\bar{R}}$ & $\mathrm{R^+}$ & Elec. & $\mathrm{R^-}$ & $\mathrm{\bar{R}}$ & $\mathrm{R^+}$ \\
\hline                       
Valence&$\mathbf{O2^{**}}$ &-0.351 &-0.0481 &0.193 &$\mathbf{AF3^{*}}$ &-0.0734 &0.0978 &0.23 &$\mathbf{F7^{**}}$ &-0.242 &0.0853 &0.324 &$\mathbf{F7^{*}}$ &-0.193 &0.0812 &0.265 \\
&$\mathbf{P8^{**}}$ &-0.276 &-0.0038 &0.233 &$\mathbf{F7^{*}}$ &-0.157 &0.0668 &0.248 &$\mathbf{FC5^{*}}$ &-0.375 &-0.0396 &0.229 &$\mathbf{T7^{*}}$ &-0.374 &0.069 &0.236 \\
&$\mathbf{T8^{**}}$ &-0.238 &-0.00455 &0.213 &$\mathbf{T7^{*}}$ &-0.322 &0.0485 &0.287 &$\mathbf{P7^{**}}$ &-0.516 &-0.0493 &0.271 &$\mathbf{O2^{*}}$ &-0.354 &-0.0277 &0.306 \\
&$\mathbf{F4^{**}}$ &-0.303 &-0.0411 &0.131 &$\mathbf{P7^{**}}$ &-0.513 &-0.039 &0.184 &$\mathbf{O1^{**}}$ &-0.385 &-0.0745 &0.33 &$\mathbf{T8^{**}}$ &-0.434 &-0.0212 &0.256 \\
&$\mathbf{F8^{*}}$ &-0.309 &-0.0223 &0.256 &$\mathbf{O1^{*}}$ &-0.369 &-0.0316 &0.256 &$\mathbf{P8^{*}}$ &-0.233 &-0.075 &0.295 &$\mathbf{FC6^{*}}$ &-0.282 &-0.0306 &0.322 \\
&$\mathbf{AF4^{*}}$ &-0.389 &0.0335 &0.373 & & & & & & & & &$\mathbf{F4^{*}}$ &-0.369 &-0.0247 &0.265 \\

\hline
Arousal&$\mathbf{AF3^{**}}$ &-0.313 &-0.00233 &0.359 &$\mathbf{AF3^{*}}$ &-0.246 &0.0197 &0.402 &$\mathbf{AF3^{*}}$ &-0.319 &-0.0366 &0.512 &$\mathbf{O1^{*}}$ &-0.325 &-0.0751 &0.0799 \\
&$\mathbf{F7^{**}}$ &-0.198 &-0.0273 &0.116 &$\mathbf{T7^{**}}$ &-0.116 &0.112 &0.296 &$\mathbf{F7^{*}}$ &-0.336 &-0.0657 &0.244 &$\mathbf{F8^{*}}$ &-0.249 &0.0273 &0.457 \\
&$\mathbf{F3^{**}}$ &-0.336 &-0.0833 &0.205 &$\mathbf{P7^{*}}$ &-0.225 &0.0506 &0.348 &$\mathbf{F3^{*}}$ &-0.331 &-0.00749 &0.44 & & & & \\
&$\mathbf{FC5^{**}}$ &-0.367 &-0.0825 &0.148 &$\mathbf{O2^{*}}$ &-0.039 &0.0832 &0.323 &$\mathbf{T7^{**}}$ &-0.0856 &0.126 &0.332 & & & & \\
&$\mathbf{P7^{**}}$ &-0.265 &-0.00849 &0.167 &$\mathbf{P8^{**}}$ &-0.293 &0.0731 &0.418 &$\mathbf{P7^{**}}$ &-0.246 &0.0723 &0.42 & & & & \\
&$\mathbf{O1^{**}}$ &-0.266 &-0.0673 &0.0999 &$\mathbf{FC6^{*}}$ &-0.146 &0.0904 &0.323 &$\mathbf{O1^{*}}$ &-0.252 &0.0372 &0.422 & & & & \\
&$\mathbf{O2^{*}}$ &-0.101 &0.011 &0.133 & & & & &$\mathbf{O2^{**}}$ &-0.154 &0.109 &0.389 & & & & \\
&$\mathbf{P8^{**}}$ &-0.357 &-0.00935 &0.34 & & & & &$\mathbf{P8^{**}}$ &-0.164 &0.104 &0.416 & & & & \\
&$\mathbf{F4^{**}}$ &-0.184 &-0.0103 &0.204 & & & & &$\mathbf{T8^{*}}$ &-0.141 &0.0693 &0.286 & & & & \\
&$\mathbf{AF4^{**}}$ &-0.24 &-0.0457 &0.0313 & & & & &$\mathbf{FC6^{*}}$ &-0.146 &0.0785 &0.394 & & & & \\
& & & & & & & & &$\mathbf{F8^{**}}$ &-0.212 &0.0471 &0.453 & & & & \\
& & & & & & & & &$\mathbf{AF4^{*}}$ &-0.285 &-0.0305 &0.283 & & & & \\
\hline
dominance	&$\mathbf{T7^{**}}$ &-0.201 &-0.0208 &0.212 &$\mathbf{O1^{*}}$ &-0.442 &-0.0423 &0.315 &$\mathbf{F7^{**}}$ &-0.489 &-0.117 &0.2 &$\mathbf{F4^{*}}$ &-0.464 &0.0407 &0.323 \\
&$\mathbf{O1^{**}}$ &-0.344 &-0.0108 &0.312 &$\mathbf{O2^{*}}$ &-0.451 &-0.0164 &0.297 &$\mathbf{F3^{*}}$ &-0.385 &-0.0615 &0.342 & & & & \\
&$\mathbf{T8^{**}}$ &-0.269 &-0.0354 &0.235 &$\mathbf{F4^{*}}$ &-0.311 &-0.0275 &0.437 &$\mathbf{O1^{*}}$ &-0.329 &-0.0656 &0.297 & & & & \\
& & & & & & & & &$\mathbf{P8^{*}}$ &-0.369 &-0.0824 &0.189 & & & & \\
& & & & & & & & &$\mathbf{T8^{*}}$ &-0.271 &-0.0748 &0.27 & & & & \\
& & & & & & & & &$\mathbf{FC6^{*}}$ &-0.31 &-0.0517 &0.236 & & & & \\
& & & & & & & & &$\mathbf{F4^{*}}$ &-0.421 &-0.0417 &0.533 & & & & \\
& & & & & & & & &$\mathbf{F8^{*}}$ &-0.34 &-0.063 &0.39 & & & & \\
& & & & & & & & &$\mathbf{AF4^{**}}$ &-0.381 &-0.101 &0.429 & & & & \\
\hline
\end{tabular}}
\end{table}


\begin{figure*}[!t]
\centering
\subfloat[ ]{\includegraphics[width=2.5in] {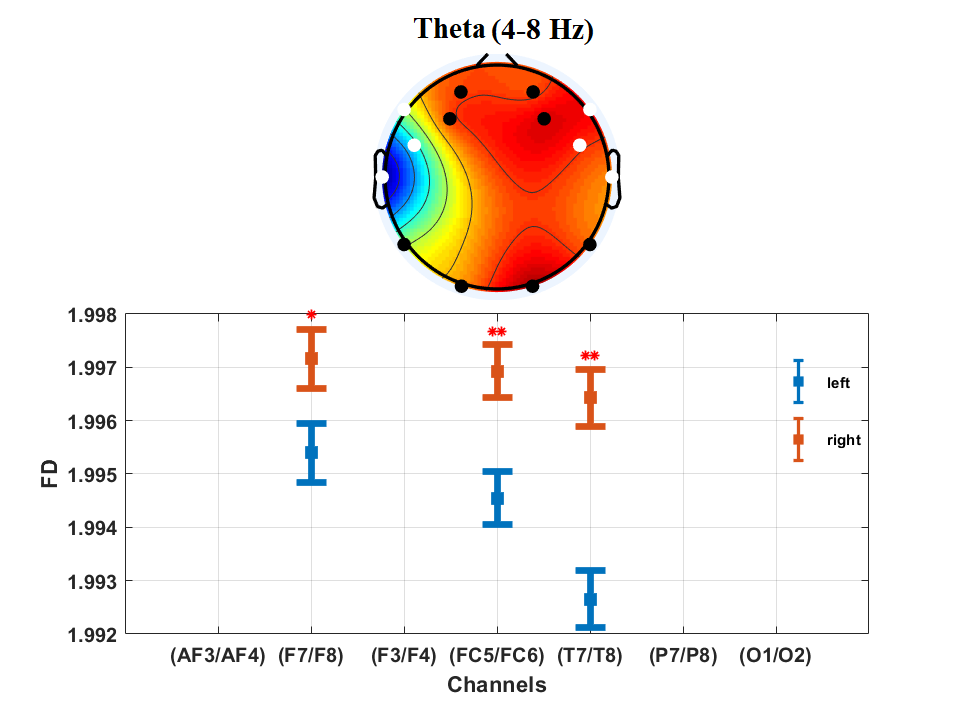}} 
\subfloat[ ]{\includegraphics[width=2.5in] {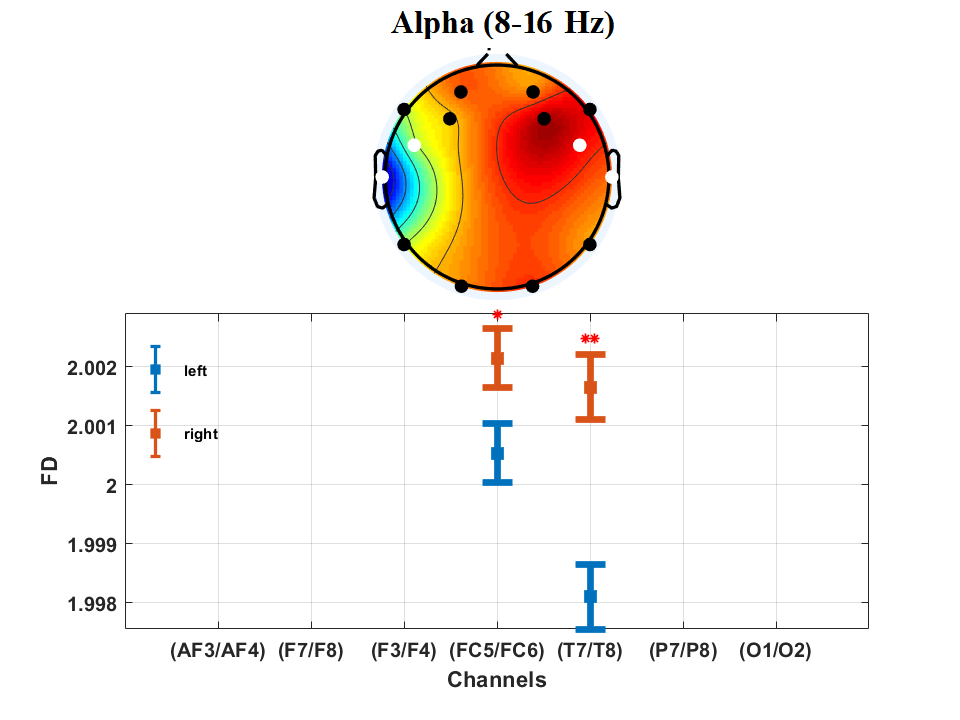}}
\\
\subfloat[ ]{\includegraphics[width=2.5in] {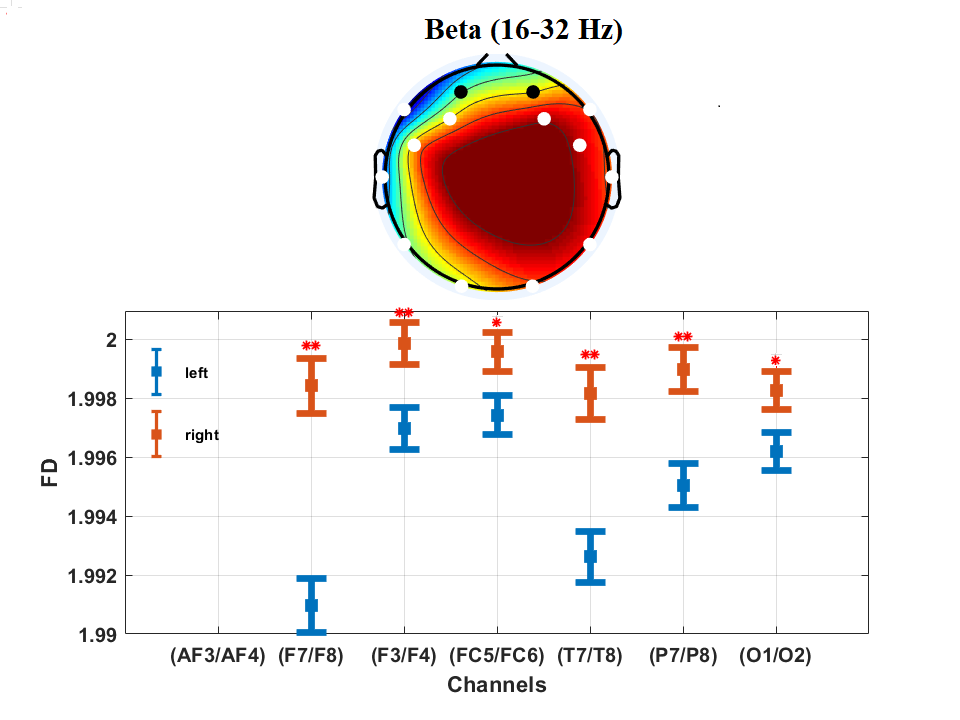}} 
\subfloat[ ]{\includegraphics[width=2.5in] {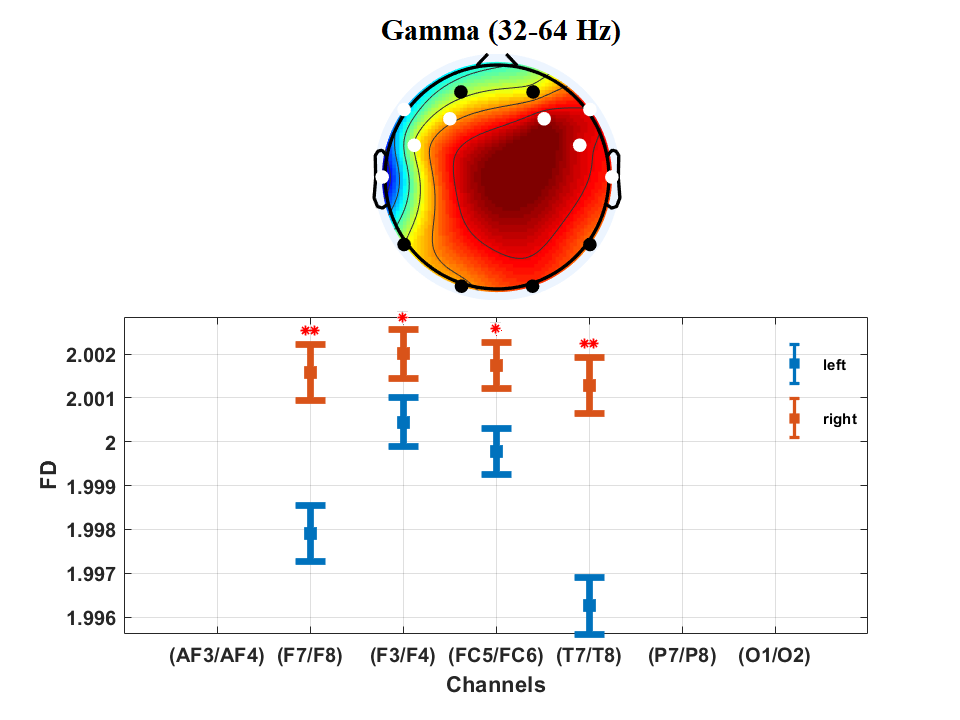}}
\caption{Comparison of the FD values between the each two symmetrical channels for the second 30 trials in term of mean$\pm$ SEM for theta (4-8 Hz), alpha (8-16 Hz), beta (16-32 Hz), and gamma (32-64 Hz), channels with significant mean differences are highlighted with white color($ \alpha = 0.05 ,^{\color{red}*}=p<0.05, ^{\color{red}**}=p<0.005$),(a) theta, (b) alpha (c) beta.}
\label{anova2}
\centering
\end{figure*}

In order to contrast fractal activities for EEG signals recorded from the behavioral performance experiment with the statistical correlation between the valence level and the fractal activities for EEG signals, a repeated-measures ANOVA is conducted by considering the variances in FD values between each two symmetrical channels in both hemispheres as between-groups variances.The individual differences were eliminated by subtracting the between participants variance form the within-group variances. Results with significant mean differences, which are depicted in Figure~\ref{anova2}, show a higher FD activities in the right hemisphere than the left hemisphere for all sub bands in the frontal and temporal areas, which according to the results in Table~\ref{electrode sub-band correlation}, suggests that the better behavioral performance is achieved, the higher valence level is experienced. 

\begin{figure*}[!t]
\centering
\subfloat[ ]{\includegraphics[width=2.5in]{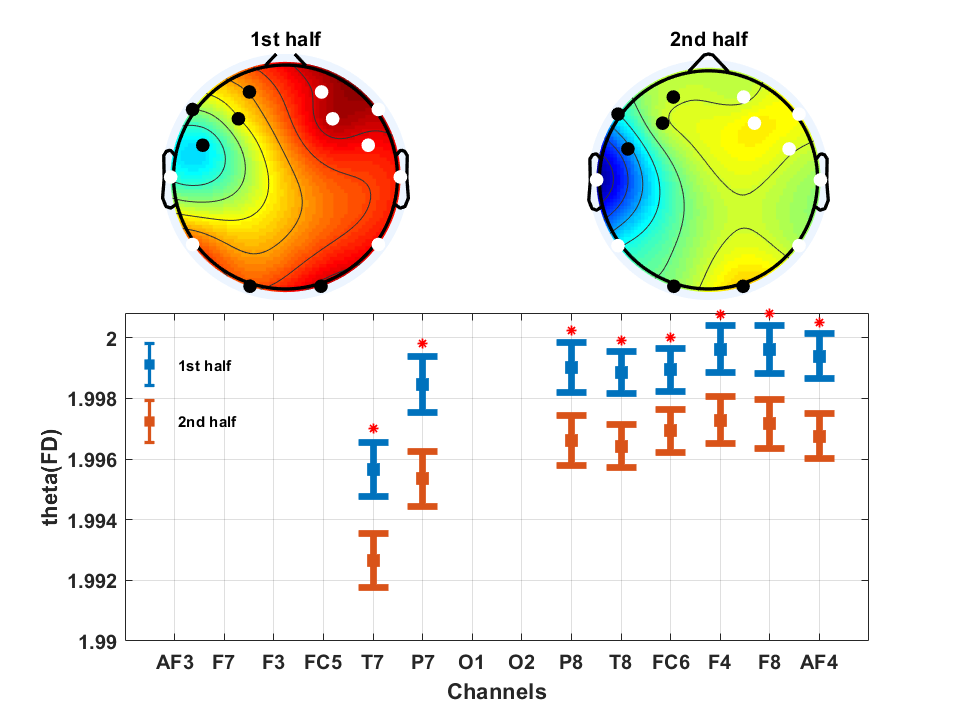}}
\subfloat[ ]{\includegraphics[width=2.5in]{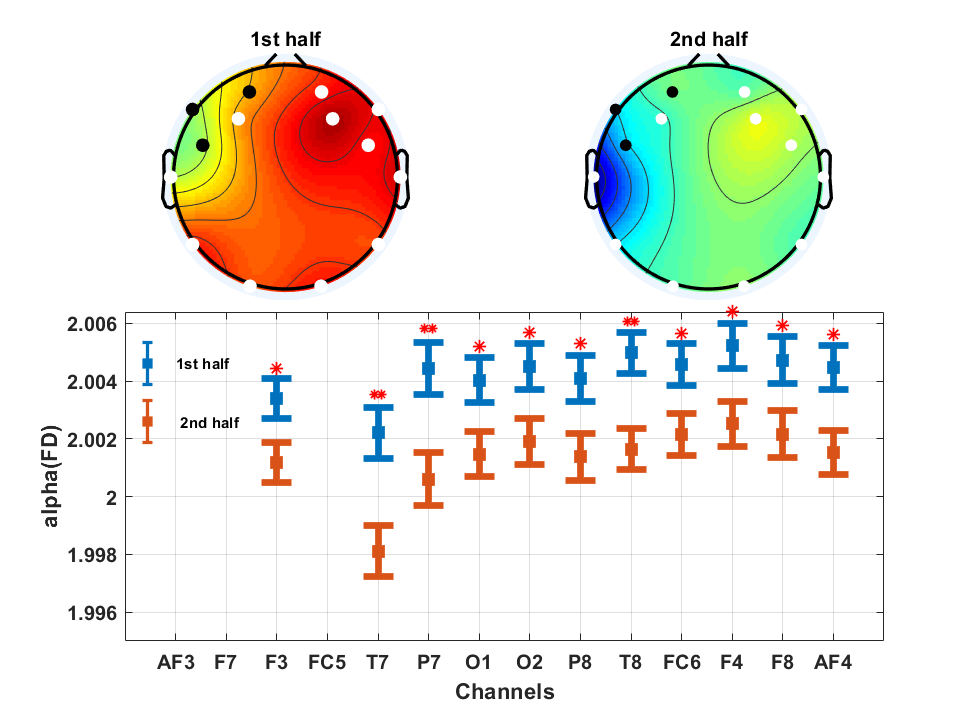}}
\\
\subfloat[ ]{\includegraphics[width=2.5in]{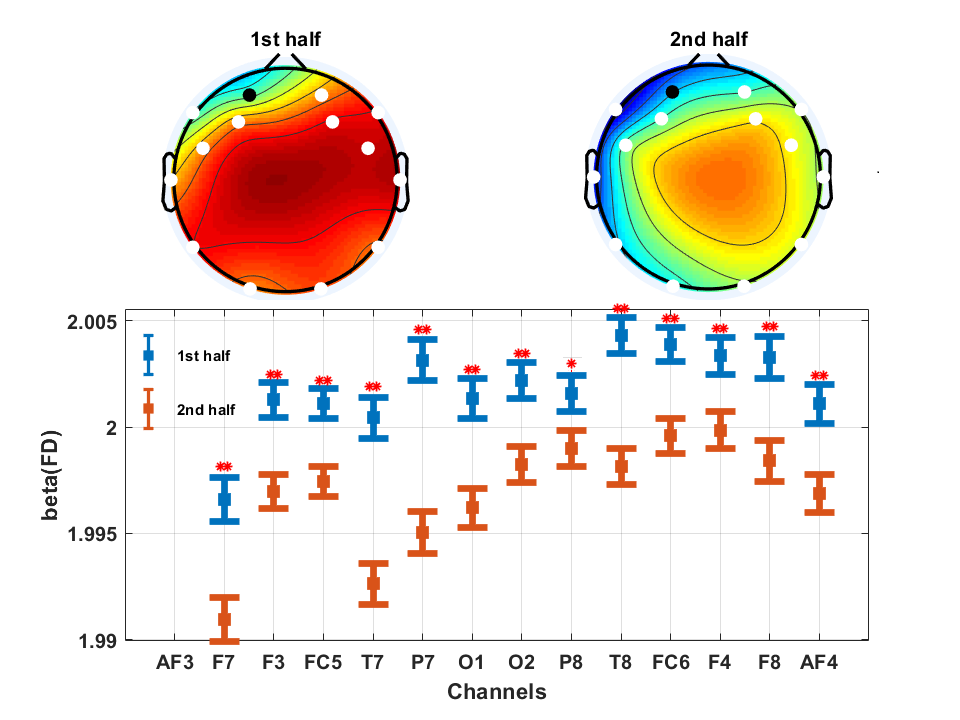}}
\subfloat[ ]{\includegraphics[width=2.5in]{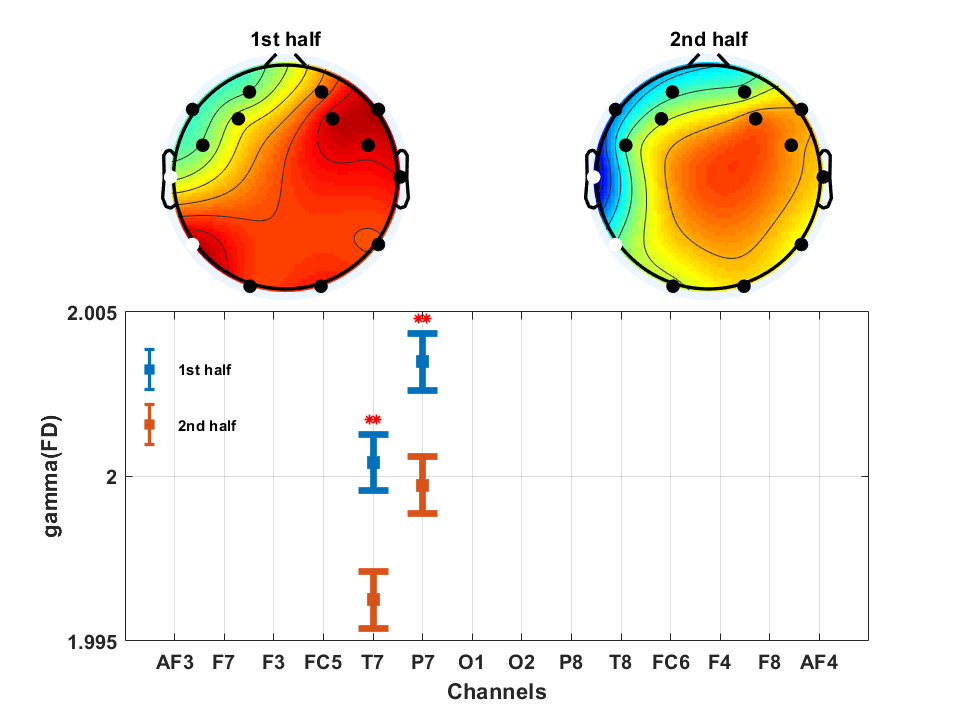}}
\caption{Comparison of the FD values between the first 30 trial and the second 30 trial in term of mean$\pm$ SEM for theta (4-8 Hz), alpha (8-16 Hz), beta (16-32 Hz), and gamma (32-64 Hz), channels with significant mean differences are highlighted with white color($ \alpha = 0.05 ,^{\color{red}*}=p<0.05, ^{\color{red}**}=p<0.005$),(a) theta, (b) alpha (c) beta.}
\label{anova1}
\centering
\end{figure*}

Another comparison with the statistical results for the arousal state is conducted by performing a repeated-measures ANOVA for each channel separately with the variance of FD values per channel between the first 30 trials and the second 30 trials as between-groups variance. The between-participants variances plus error variances were considered as within-group variance. The individual differences were eliminated by subtracting the between participants variance form the within-group variances. Results with significant means differences are shown in Figure~\ref{anova1} for theta, beta, alpha, and gamma sub bands.
The results show a significant drop in FD values from the first 30 trials to the second 30 trials for all bands in the temporal, parietal, and occipital. A comparison between these results and the results for the arousal state in Table~\ref{electrode sub-band correlation}, indicates that the degradation in the behavioral performance is associated with a decrease in the arousal level.

\subsubsection{PAD states Gaussian Process} \label{PAD states Gaussian Process} 
The datasets for EEG-based FD and EEG bands-based FD are used to infer EEG-based GP regression model and EEG bands-based GP regression model, respectively. As it is discussed in Section~\ref{Inference of Behavioral Performance}, the covariance function for each model is obtained by training DBN using an unlabeled dataset denoted as pre-training dataset which is composed of the EEG-based measurements collected during the performance induction experiment. Then, the resulting DBN is used to initialize a covariance function of Gaussian process to perform a supervised learning using a labeled dataset denoted as fine-tuning dataset, which is composed of EEG-based measurements and their corresponding PAD states labels collected from the emotion elicitation experiment. Hence, two types of pre training datasets are obtained, namely, EEG-based and EEG bands-based training sets and two types of fine-tuning datasets are obtained, namely, EEG-based and EEG bands-based fine-tuning sets. 

\begin{figure*}[!t]
\centering
\subfloat[]
{
\includegraphics[width=3in]{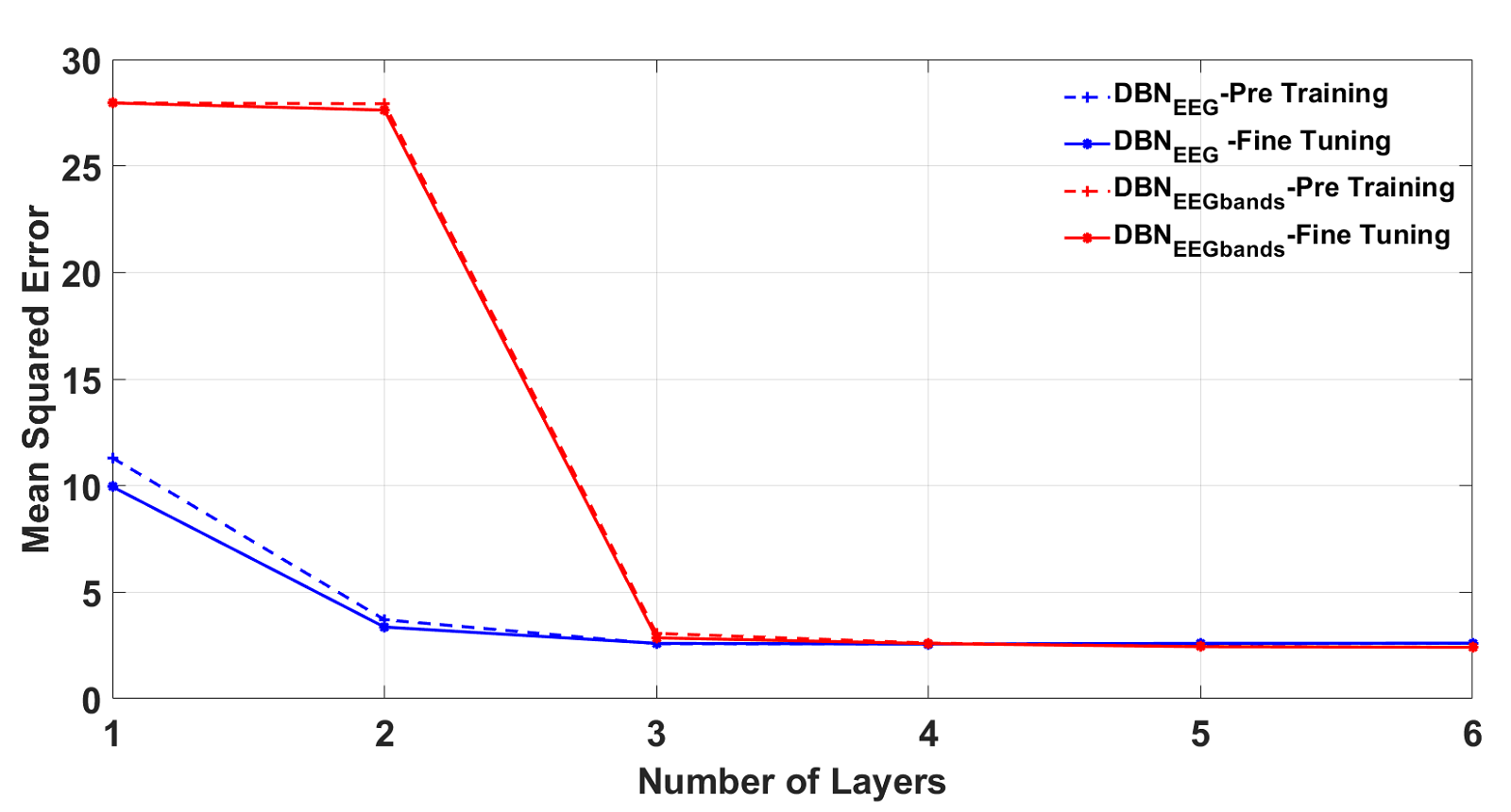}
\label{6:MSE}
}
\\
\subfloat[]
{
\includegraphics[width=3in]{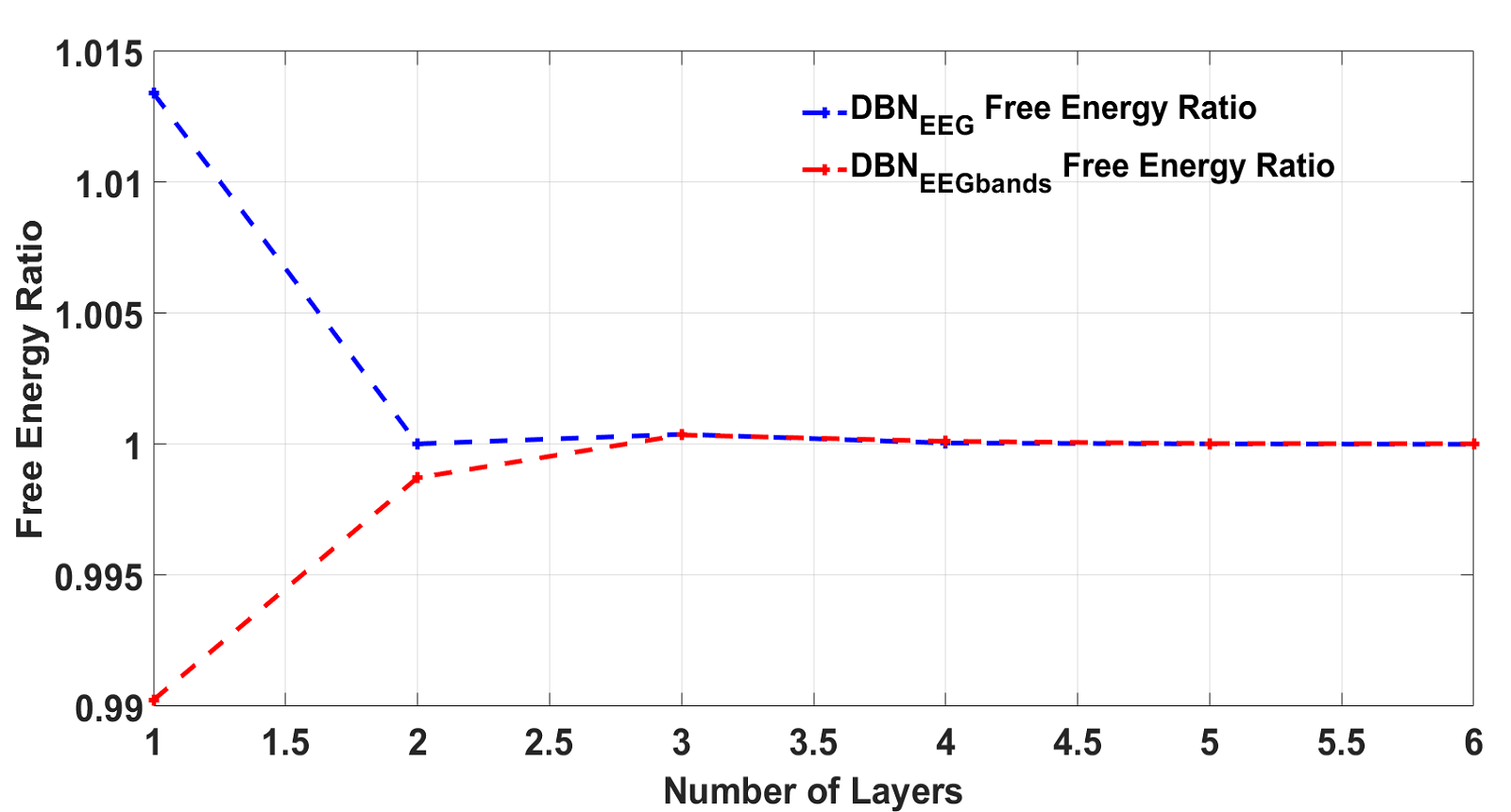}
\label{7:energy ratio}
}
\caption{(a) The mean square error for different DBN with different number of layers during pre-training and fine tuning stages and (b)energy ratio for different DBN with different number of layers}
\centering
\end{figure*}

For the DBN of each GP model, 20 and 80 hidden units are used for the DBN of EEG-based and EEG band-based GP models, respectively, where the number of hidden units is determined based on the number and the dimension of the training samples ~\cite{hinton2012practical}. The EEG-based and EEG bands-based pre-training datasets are used for the greedy layer-by-layer pre-training of the DBN~\cite{hinton2006fast}. For each layer, single-step Contrastive Divergence (1-CD) algorithm is used for the training. The learning rate for the training of first hidden layer was set to $0.01$, and for the higher hidden layers was set to $0.001$. The weight decay and momentum parameters were set to $0.0002$ and $0.1$ respectively, for the training of all layers. The ratio between the free energy of the EEG-based pre-training dataset and EEG-based validating dataset were calculated to monitor the overfitting~\cite{hinton2012practical}. The evolution of free energy ratio during the greedy layer-wise pre-training is shown in Figure~\ref{7:energy ratio} by the dashed blue line. 
After adding each layer, the EEG-based and EEG bands-based fine-tuning datasets were used to update the weights of each DBN using backpropagation~\cite{haykin2009neural}. The learning rate was set to $10^{-1}$ and annealed linearly to $10^{-5}$. The momentum and weight decay parameters were set to $0.1$. Fine tuning process decreases the mean square errors for the two models with all numbers of layers as it can be see in Figure~~\ref{6:MSE}, in which the solid red line and solid blue line represent the mean square error of the EEG-based and EEG sub bands-based GP models, respectively. 

The mean square error for EEG-bands GP model showed no significant improvement after the fifth layer as it is shown in Figure~~\ref{6:MSE} by the dashed red line, resulting in a DBN with architecture of $(56-80-80-80-80)$. No significant over fitting was recorded  during the pre-training stage as it can be seen in Figure~~\ref{7:energy ratio} by the dotted red line. Similarly, the EEG-based GP model stopped showing any significant improvement in term of mean square error after the fifth layer, resulting in a DBN with architecture of $(14-20-20-20-20)$. The improvement in the mean squared error over the course of layers adding is shown in Figure~~\ref{6:MSE} by the dashed blue line. The GP model based on EEG bands required DBN with more layers than that required for EEG-based GP model before settling the mean square error. However, it showed superior accuracy relative to the EEG-based GP. This result is consistent with the reported significant oscillatory correlation between EEG signal and PAD states~\cite{muhl2014survey}. Most of the previous efforts have exploited the power spectrum as a feature to identify different PAD states. Power spectrum is no stranger to the fractal geometry since it has been used extensively to extract indices for signal complexity~\cite{higuchi1988approach}. This is done by calculating the power spectrum at different frequencies $S(f_i), i=1,2,\dots.$, and then using the power law $S(f_i)  \propto f_{i}^{\alpha}$ to extract $\alpha$ which represents the irregularity in time series signals. The use of power spectrum to scale signal complexity in addition to the reported significant correlation between emotions and power spectrum of certain EEG sub bands, points out the ability to extract more distinct emotion-relevant information from the EEG spectrum-based complexity indices, which is supported by the results in this subsection.

In order to examine the advantage of learning Gaussian kernel using DBN, the EEG-based and EEG bands-based GP models were fitted using spherical Gaussian (SG) kernel. A comparison between mean square errors resulted from using the two kinds of Gaussian kernels showed that learning Gaussian kernel with DBN improved the accuracy significantly by decreasing the mean square error for the validation set, as it is shown in Table~\ref{MSE table}.

\begin{table}
\caption {mean square error for PAD states GP models}
\label{MSE table}
\centering
\begin{tabular}{|c|c|c|c|}
\hline
\multicolumn{1}{|c|}{\textbf{model~type}}   &\multicolumn{1}{|c|}{\textbf{GP with DBN kernel }}   &\multicolumn{1}{|c|}{\textbf{GP with SG Kernel}}\\
\hline
EEG-based            &$2.5654$   &$26.2982$ \\
\hline
EEG bands-based       &$2.4438$   &$27.9547$\\
\hline
\end{tabular}
\end{table}

\subsection{Behavioral Performance Inference}\label{Behavioral Performance Inference}
In this part, the emotional features of EEG signals represented by the PAD states are used to estimate behavioral performance. The PAD states of the participants are estimated using PAD states Gaussian Process based on the EEG signals collected in the behavioral performance induction experiment. Similar to the PAD states inference, two types of emotional features, which are the EEG-based PAD states and EEG bands-based PAD states, are used to obtain two GP models. The two GP models are compared in term of prediction accuracy. 

\subsubsection{Emotional Features for EEG}
The EEG data collected during the behavioral performance induction were processed using the procedure presented in section~\ref{fractal feature} to obtain EEG-based FD dataset and EEG bands-based FD dataset. Then the two datasets are used to obtain EEG-based PAD dataset and EEG bands-based dataset using the EEG-based GP model and EEG bands-based GP model, respectively.

\begin{figure*}[!t]
\centering
\subfloat[ ]
{
\includegraphics[width=3in]{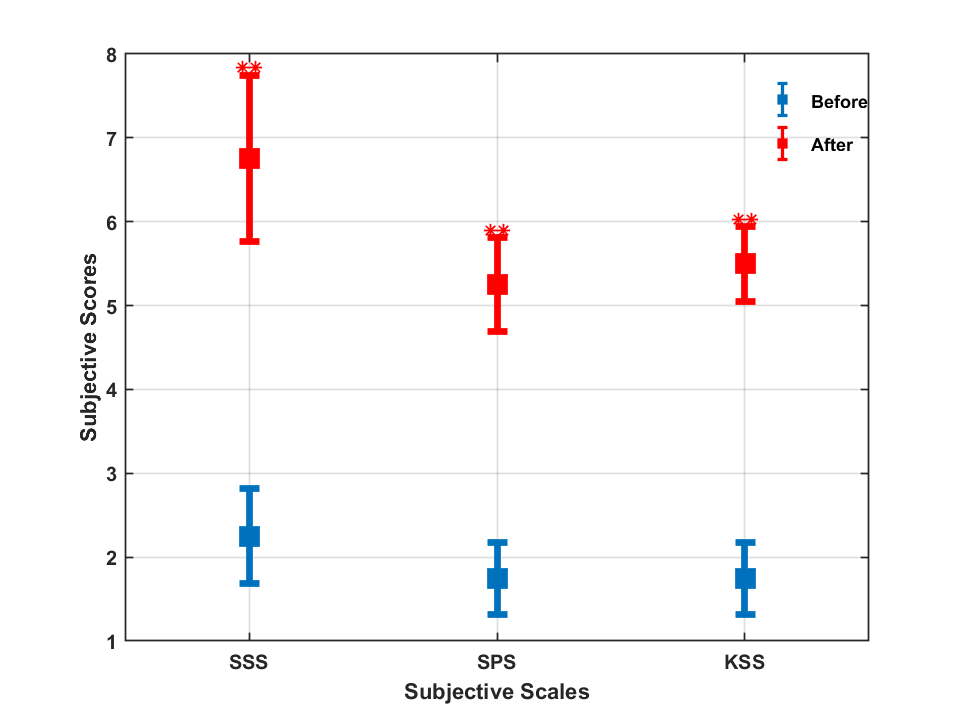}
\label{8:fatigue}
}
\\
\subfloat[ ]
{
\includegraphics[width=3in]{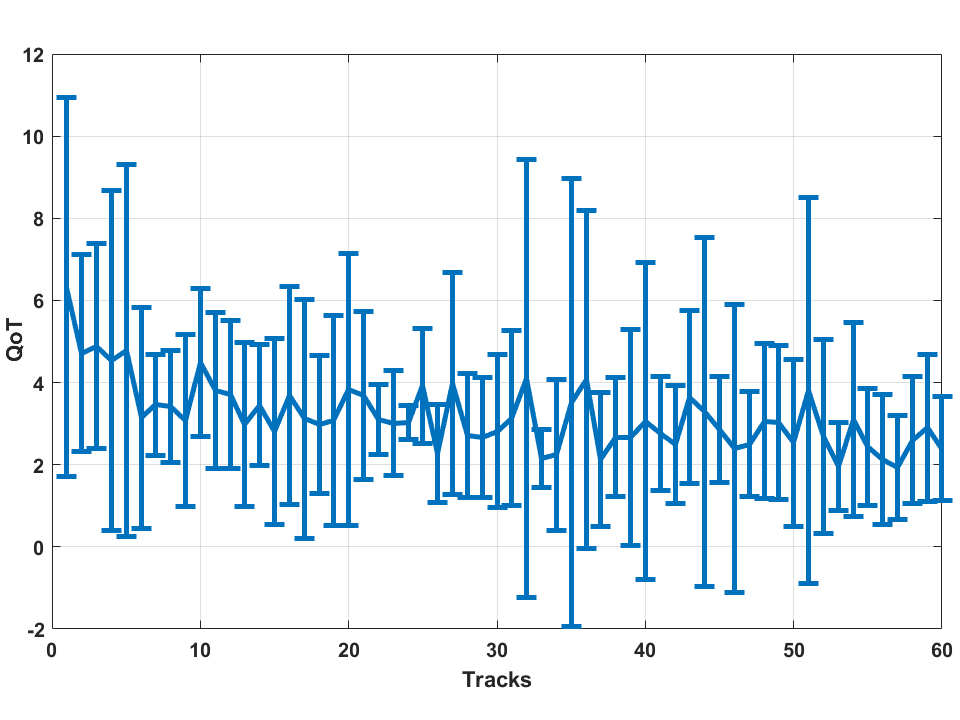}
\label{8:QoT}
}
\caption{(a)The self reported fatigue level based on SSS, KPS, and KSS scales. mean$\pm$ SEM, $(^{\color{red}*}=p<0.05, ^{\color{red}**}=p<0.005)$ and (b)The QoT evaluation over the course of performed trials(mean$\pm$ SEM)}
\centering
\end{figure*}

For statistical analysis, a comparison between self reported fatigue level in the SSS, SPS, and KSS questionnaires before and after the experiment shows a significant increase in the fatigue ($p<0.005$) as it is shown Figure~\ref{8:fatigue}. Moreover, the single-tailed (left) Pearson correlation for each participant shows a significant negative correlations ($r<-0.33,p<0.001$) between the QoT value and the number of accomplished trials. Hence, the results show that the increasing fatigue level negatively impacts on the QoT value,i.e., as the fatigue level increases the QoT values decreases. The averaged QoT values for all participant over the course of trials is shown in Figure~\ref{8:QoT}. To investigate the relationship between the PAD states and behavioral performance, we calculate the Spearman correlation coefficient $\mathrm{R}$ between each of PAD states per trial and QoT per trial for all participants. The results which are shown in table~\ref{emotion_corr} show significant positive correlations between QoT and each of valence, arousal, and dominance which is compatible with our earlier results that show a significant positive correlation ($p<0.05$) between valence and dominance, with QoT values \cite{alfatlawi2015eeg}. According to these results, low quality of behavioral performance has a negative influence on valence state. This negative correlation is consistent with the findings in~\cite{dittner2004assessment}, which show that people with depression are more likely to report high fatigue levels. 
The repetitive nature of tasks allows the participants to acquire the necessary skills to optimize their performance. After that, the arousal state degrades as the participants keep performing tasks with the same level of complexity which explains the positive correlation between the QoT level and arousal. This result coincides with flow theory which states that skills level has a negative influence on the arousal state because as the former increases, the targeted task becomes less challenging \cite{seligman2014positive}.
Dominance can be defined as the feeling of ability to affect and control versus being affected and controlled by the surrounding environment\cite{russell1977evidence}. By mapping this definition to the task environment in the behavioral performance induction experiment, it can be seen that as level of QoT decreases, participant's control level on the manipulator degrades. The awareness of losing control, through the visual feedback, is reflected negatively on the dominance state which leads to the negative correlation. 
\begin{table}[b!]
\caption {PAD states with significant correlation ($ ^*=p<0.05, ^{**}=p<0.005$)}
\label{emotion_corr}
\centering
\begin{tabular}{c c c}
\hline
\hline
\textbf{Emotional State}& $\mathrm{R}$ \\
\hline
$\mathbf{Valence^*}$      &$0.11$ \\
\hline
$\mathbf{Arousal^*}$      &$0.094$  \\
\hline
$\mathbf{Dominance^*}$  &$0.096$ \\
\hline
\end{tabular}
\end{table}

\subsubsection{Behavioral Performance Gaussian Process }
The EEG-based PAD states and EEG bands-based PAD states datasets are used to obtain two GP models, namely, EEG-based GP behavioral model and EEG bands-based behavioral model. We perform a 5-folds repeated Cross Validation (CV) for model selection~\cite{haykin2009neural} and grid-search to obtain the variance $\alpha_g \in (0.01,0.1)$ and length scale $\beta_g \in (1,3)$ for the covariance function in equation~\ref{performance covariance function}. The mean square error results for both GP models, which are presented in Table~\ref{MSE performance table}, shows the impact of utilizing the spectral features in improving the prediction accuracy for behavioral performance.  
\begin{table}
\caption {mean square error for GP behavioral models}
\label{MSE performance table}
\centering
\begin{tabular}{|c|c|c|c|}
\hline
\multicolumn{1}{|c|}{\textbf{model~type}}   &\multicolumn{1}{|c|}{\textbf{GP with DBN kernel }} \\
\hline
EEG-based            &$2.2191$  \\
\hline
EEG bands-based       &$1.3931$ \\
\hline
\end{tabular}
\end{table}


\subsection{Induced Behavioral Performance Closed Loop System}
\begin{figure}[!t]
\centering
\subfloat[ ]{\includegraphics[width=2.5in] {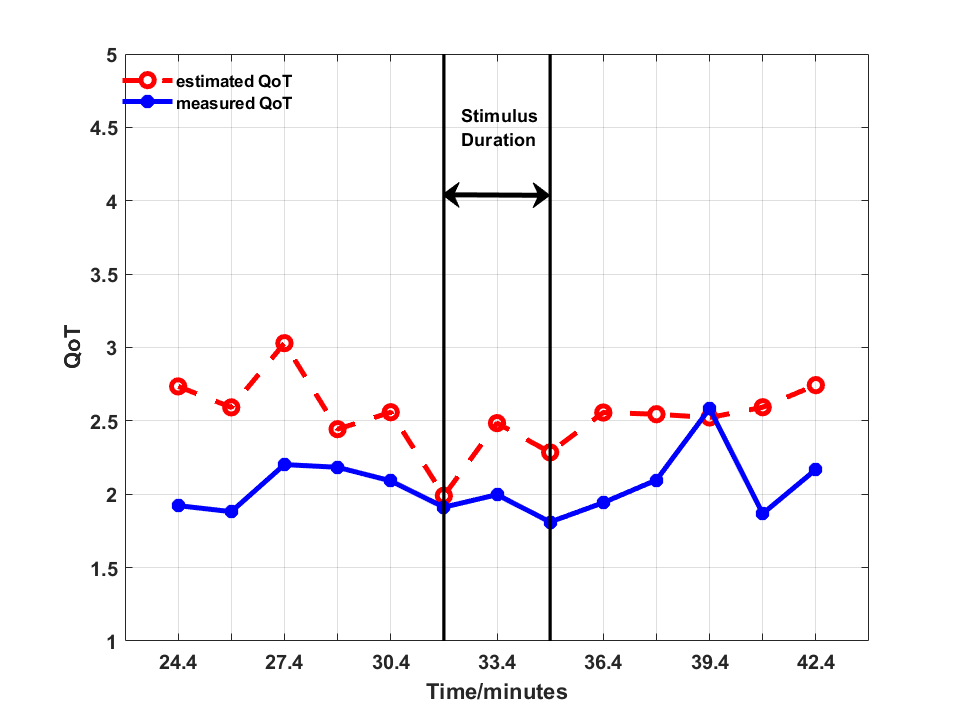}} 
\subfloat[ ]{\includegraphics[width=2.5in] {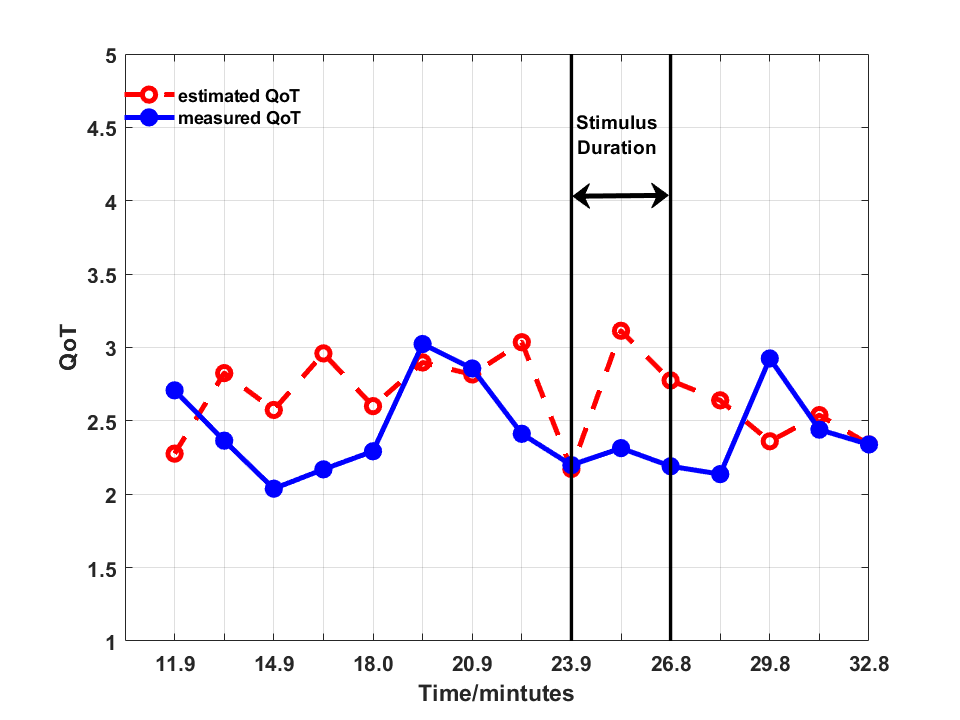}} 
\\
\subfloat[ ]{\includegraphics[width=2.5in] {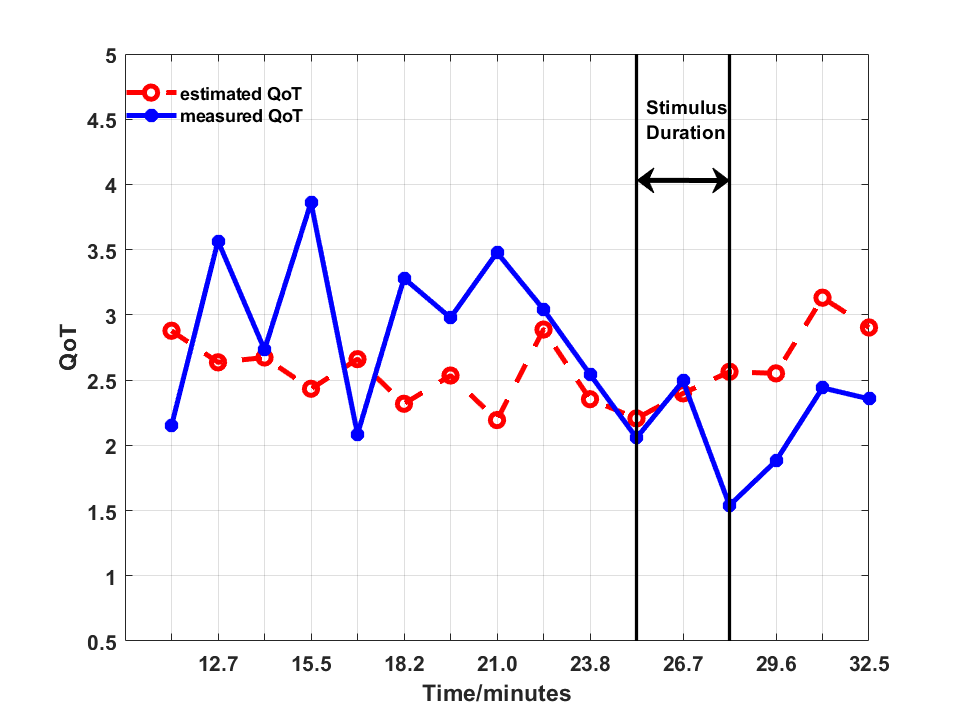}} 
\caption{Estimated QoT and performance-indices based QoT values before and after stimulus instant represented by the solid black vertical line. for (a) participant no.1, (b) participant no.2, (c) participant no. 3}
\label{3sub_qot}
\centering
\end{figure}

The results of the implementation of the closed loop system presented in Figure~\ref{CL_performance} using the GP models obtained in sections~\ref{PAD states Inference} and~\ref{Behavioral Performance Inference} are shown in~Figure~\ref{3sub_qot} which represent the estimated QoT values and measured QoT values before, after, and during the stimulus which is marked by the two solid vertical black line. For participant 1,  improvement in performance is shown after the end of the stimulus. This instantaneous improvement cannot be seen for participant 2 and participant 3. One possible explanation of improvement in performance of participant 1 is that s/he was familiar with the played audio stimulus which enable him to make use of the elicited positive emotion to improve his performance during the following time. Unfortunately, this was not the case with participant 2 and participant 3 who were not familiar with the played audio stimulus which made them distracted form the task and focusing more on the audio stimulus, and led to degradation in performance. Negative emotions were generated as a result of this degradation which encountered the positive emotions boosted by the audio stimulus. The interplay between the positive emotions (resulted form audio stimulus) and negative emotions (resulted from degradation in performance) ended after the end of the audio stimulus for the two cases in which a relative improvement in performance is observed. According to the results in this experiment, non-familiarness of participants with the audio stimulus represents a major flaw in the design of the experiment. Also the cultural background of participants should be considered when choosing audio stimuli which impose a serious challenge to this approach because of the shortage of credible emotionally provocative audio stimuli with cultural variety. 

\section{Conclusions}\label{conclusion}
The bidirectional interaction between emotions and behavioral performance can be leveraged to ensure the public safety and optimize the outcome for HRI systems. In this work, EEG-based features, PAD states, and behavioral performance are used to identify the states of a controllable and observable model for the behavioral performance. The identified model defines the EEG-based features as the output, from which the PAD states and the concurrent behavioral performance can be observed. The PAD states serves as controllable states of the model that can be influenced through an external emotional stimulus as an input. The experimental results based on the proposed model show significant positive correlations between $QoT$ values and PAD states, which are consistent with reported positive correlations between performance and positive emotions in several studies. A major advantage of the proposed model is that it facilitates the design of a closed loop system which is influenced by external emotional stimulus to maintain the behavioral performance above the safety level. The effectiveness for the proposed closed loop system is manifested in the significant improvement in behavioral performance observed after the delivery of the audio stimulus. 
A drawback for the proposed closed loop system is the uncertainty in the level of elicited PAD states associated with each audio stimulus. This uncertainty results from the dependency of the musical preference for human operators, and on the cultural background, age, personality and many other factors.

\clearpage
\bibliography{main}

\begin{thebibliography}{10}
\expandafter\ifx\csname url\endcsname\relax
  \def\url#1{\texttt{#1}}\fi
\expandafter\ifx\csname urlprefix\endcsname\relax\def\urlprefix{URL }\fi
\expandafter\ifx\csname href\endcsname\relax
  \def\href#1#2{#2} \def\path#1{#1}\fi

\bibitem{hagele2016robots}
M.~Hagele, Robots conquer the world [turning point], IEEE Robotics \&
  Automation Magazine 23~(1) (2016) 120--118.

\bibitem{cummings2014man}
M.~M. Cummings, Man versus machine or man+ machine?, IEEE Intelligent Systems
  29~(5) (2014) 62--69.

\bibitem{muro2015robots}
M.~Muro, S.~Andes, Robots seem to be improving productivity, not costing jobs,
  Harvard business review.

\bibitem{graetz2015estimating}
G.~Graetz, G.~Michaels, Estimating the impact of robots on productivity and
  employment, Center for Economic Performance.

\bibitem{engelberger2012robotics}
J.~F. Engelberger, Robotics in Practice: Management and Applications of
  Industrial Rrobots, Springer Science \& Business Media, 2012.

\bibitem{national2007human}
N.~R. Council, et~al., Human-System Integration in the System Development
  Process: A New Look, National Academies Press, 2007.

\bibitem{mitler1988catastrophes}
M.~M. Mitler, M.~A. Carskadon, C.~A. Czeisier, W.~C. Dement, D.~F. Dinges,
  R.~C. Graeber, Catastrophes, sleep, and public policy: consensus report,
  Sleep 11~(1) (1988) 100--109.

\bibitem{reiner2012insidious}
B.~I. Reiner, E.~Krupinski, The insidious problem of fatigue in medical imaging
  practice, Journal of Digital Imaging 25~(1) (2012) 3--6.

\bibitem{wheaton2014drowsy}
A.~G. Wheaton, R.~A. Shults, D.~P. Chapman, E.~S. Ford, J.~B. Croft, Drowsy
  driving and risk behaviors—10 states and puerto rico, 2011--2012, MMWR.
  Morbidity and Mortality Weekly Report 63~(26) (2014) 557.

\bibitem{marathe2017confidence}
A.~R. Marathe, J.~R. McDaniel, S.~M. Gordon, K.~McDowell, Confidence-based
  state estimation: A novel tool for test and evaluation of human-systems, in:
  Advances in Human Factors in Robots and Unmanned Systems, Springer, 2017, pp.
  291--303.

\bibitem{li2008measurement}
H.-C.~O. Li, J.~Seo, K.~Kham, S.~Lee, Measurement of 3{D} visual fatigue using
  event-related potential ({ERP}): 3{D} oddball paradigm, in: 2008 3DTV
  Conference: The True Vision-Capture, Transmission and Display of 3D Video,
  IEEE, 2008, pp. 213--216.

\bibitem{murata2005evaluation}
A.~Murata, A.~Uetake, Y.~Takasawa, Evaluation of mental fatigue using feature
  parameter extracted from event-related potential, International Journal of
  Industrial Ergonomics 35~(8) (2005) 761--770.

\bibitem{liu2018eeg}
Y.~Liu, Z.~Lan, H.~H.~G. Khoo, K.~H.~H. Li, O.~Sourina, W.~Mueller-Wittig,
  {EEG}-based evaluation of mental fatigue using machine learning algorithms,
  in: 2018 International Conference on Cyberworlds ({CW}), IEEE, 2018, pp.
  276--279.

\bibitem{hu2017comparison}
J.~Hu, Comparison of different features and classifiers for driver fatigue
  detection based on a single eeg channel, Computational and Mathematical
  Methods in Medicine 2017.

\bibitem{michail2008eeg}
E.~Michail, A.~Kokonozi, I.~Chouvarda, N.~Maglaveras, {EEG} and {HRV} markers
  of sleepiness and loss of control during car driving, in: 2008 30th Annual
  International Conference of the IEEE Engineering in Medicine and Biology
  Society, IEEE, 2008, pp. 2566--2569.

\bibitem{lin2010tonic}
C.-T. Lin, K.-C. Huang, C.-F. Chao, J.-A. Chen, T.-W. Chiu, L.-W. Ko, T.-P.
  Jung, Tonic and phasic {EEG} and behavioral changes induced by arousing
  feedback, NeuroImage 52~(2) (2010) 633--642.

\bibitem{huang2016eeg}
K.-C. Huang, T.-Y. Huang, C.-H. Chuang, J.-T. King, Y.-K. Wang, C.-T. Lin,
  T.-P. Jung, An {EEG}-based fatigue detection and mitigation system,
  International Journal of Neural Systems 26~(04) (2016) 1650018.

\bibitem{cai2011modeling}
H.~Cai, Y.~Lin, Modeling of operators' emotion and task performance in a
  virtual driving environment, International Journal of Human-Computer Studies
  69~(9) (2011) 571--586.

\bibitem{pekrun2009achievement}
R.~Pekrun, A.~J. Elliot, M.~A. Maier, Achievement goals and achievement
  emotions: Testing a model of their joint relations with academic
  performance., Journal of Educational Psychology 101~(1) (2009) 115.

\bibitem{seve2007performance}
C.~S{\`e}ve, L.~Ria, G.~Poizat, J.~Saury, M.~Durand, Performance-induced
  emotions experienced during high-stakes table tennis matches, Psychology of
  Sport and Exercise 8~(1) (2007) 25--46.

\bibitem{isen1993positive}
A.~M. Isen, An influence of positive affect on decision making in complex
  situations: Theoretical issues with practical implications, Journal of
  Consumer Psychology 11~(2) (2001) 75--85.

\bibitem{murphy2012oxford}
S.~Murphy, The Oxford Handbook of Sport and Performance Psychology, Oxford
  University Press, 2012.

\bibitem{fisher2004within}
C.~D. Fisher, C.~S. Noble, A within-person examination of correlates of
  performance and emotions while working, Human Performance 17~(2) (2004)
  145--168.

\bibitem{jia2014quality}
Y.~Jia, N.~Xi, S.~Liu, Y.~Wang, X.~Li, S.~Bi, Quality of teleoperator adaptive
  control for telerobotic operations, The International Journal of Robotics
  Research 33~(14) (2014) 1765--1781.

\bibitem{plutchik2003emotions}
R.~Plutchik, Emotions and life: Perspectives From Psychology, Biology, and
  Evolution., American Psychological Association, 2003.

\bibitem{russell1977evidence}
J.~A. Russell, A.~Mehrabian, Evidence for a three-factor theory of emotions,
  Journal of Research in Personality 11~(3) (1977) 273--294.

\bibitem{russell1980circumplex}
J.~A. Russell, A circumplex model of affect., Journal of Personality and Social
  Psychology 39~(6) (1980) 1161.

\bibitem{muhl2014survey}
C.~M{\"u}hl, B.~Allison, A.~Nijholt, G.~Chanel, A survey of affective brain
  computer interfaces: principles, state-of-the-art, and challenges,
  Brain-Computer Interfaces 1~(2) (2014) 66--84.

\bibitem{butler2007saddlepoint}
R.~W. Butler, Saddlepoint Approximations with Applications, Vol.~22, Cambridge
  University Press, 2007.

\bibitem{hinton2008using}
G.~E. Hinton, R.~R. Salakhutdinov, Using deep belief nets to learn covariance
  kernels for gaussian processes, in: Advances in Neural Information Processing
  Systems, 2008, pp. 1249--1256.

\bibitem{haykin2009neural}
S.~Haykin, Neural Networks and Learning Machines, Pearson, 2009.

\bibitem{lee1990fuzzy}
C.-C. Lee, Fuzzy logic in control systems: fuzzy logic controller. i, IEEE
  Transactions on systems, man, and cybernetics 20~(2) (1990) 404--418.

\bibitem{lang1997international}
P.~J. Lang, M.~M. Bradley, B.~N. Cuthbert, et~al., International affective
  picture system ({IAPS}): Technical manual and affective ratings, NIMH Center
  for the Study of Emotion and Attention 1 (1997) 39--58.

\bibitem{koelstra2011deap}
S.~Koelstra, C.~Muhl, M.~Soleymani, J.-S. Lee, A.~Yazdani, T.~Ebrahimi, T.~Pun,
  A.~Nijholt, I.~Patras, {DEAP}: A database for emotion analysis; using
  physiological signals, IEEE Transactions on Affective Computing 3~(1) (2011)
  18--31.

\bibitem{lane2011physiological}
A.~M. Lane, M.~G. Wilson, G.~P. Whyte, R.~Shave, Physiological correlates of
  emotion-regulation during prolonged cycling performance, Applied
  Psychophysiology and Biofeedback 36~(3) (2011) 181--184.

\bibitem{adeli2007wavelet}
H.~Adeli, S.~Ghosh-Dastidar, N.~Dadmehr, A wavelet-chaos methodology for
  analysis of eegs and eeg subbands to detect seizure and epilepsy, IEEE
  Transactions on Biomedical Engineering 54~(2) (2007) 205--211.

\bibitem{liu2013eeg}
Y.~Liu, O.~Sourina, {EEG} databases for emotion recognition, in: 2013
  International Conference on Cyberworlds, IEEE, 2013, pp. 302--309.

\bibitem{gomez2006automatic}
G.~G{\'o}mez-Herrero, W.~De~Clercq, H.~Anwar, O.~Kara, K.~Egiazarian,
  S.~Van~Huffel, W.~Van~Paesschen, Automatic removal of ocular artifacts in the
  {EEG} without an {EOG} reference channel, in: Proceedings of the 7th Nordic
  Signal Processing Symposium-NORSIG 2006, IEEE, 2006, pp. 130--133.

\bibitem{delorme2004eeglab}
A.~Delorme, S.~Makeig, {EEGLAB}: an open source toolbox for analysis of
  single-trial {EEG} dynamics including independent component analysis, Journal
  of Neuroscience Methods 134~(1) (2004) 9--21.

\bibitem{adeli2003analysis}
H.~Adeli, Z.~Zhou, N.~Dadmehr, Analysis of {EEG} records in an epileptic
  patient using wavelet transform, Journal of Neuroscience Methods 123~(1)
  (2003) 69--87.

\bibitem{higuchi1988approach}
T.~Higuchi, Approach to an irregular time series on the basis of the fractal
  theory, Physica D: Nonlinear Phenomena 31~(2) (1988) 277--283.

\bibitem{bailey1998combining}
T.~L. Bailey, M.~Gribskov, Combining evidence using p-values: application to
  sequence homology searches., Bioinformatics (Oxford, England) 14~(1) (1998)
  48--54.

\bibitem{pfurtscheller2001functional}
G.~Pfurtscheller, Functional brain imaging based on {ERD}/{ERS}, Vision
  Research 41~(10-11) (2001) 1257--1260.

\bibitem{hinton2012practical}
G.~E. Hinton, A practical guide to training restricted boltzmann machines, in:
  Neural Networks: Tricks of the Trade, Springer, 2012, pp. 599--619.

\bibitem{hinton2006fast}
G.~E. Hinton, S.~Osindero, Y.-W. Teh, A fast learning algorithm for deep belief
  nets, Neural Computation 18~(7) (2006) 1527--1554.

\bibitem{alfatlawi2015eeg}
M.~Alfatlawi, Y.~Jia, N.~Xi, {EEG}-based quality of teleoperator identification
  using emotional states model, in: 2015 IEEE International Conference on
  Robotics and Biomimetics (ROBIO), IEEE, 2015, pp. 470--475.

\bibitem{dittner2004assessment}
A.~J. Dittner, S.~C. Wessely, R.~G. Brown, The assessment of fatigue: a
  practical guide for clinicians and researchers, Journal of Psychosomatic
  Research 56~(2) (2004) 157--170.

\bibitem{seligman2014positive}
M.~E. Seligman, M.~Csikszentmihalyi, Positive psychology: An introduction, in:
  Flow and the Foundations of Positive Psychology, Springer, 2014, pp.
  279--298.

\end{thebibliography}

\end{document}